%% file: ms-mdwarf-radii.tex
\pdfoutput=1
\documentclass[fleqn, usenatbib]{mnras}
\usepackage{graphicx}	
\usepackage{amsmath}	
\usepackage{amssymb}	
\usepackage[version=3]{mhchem}
\usepackage{newtxtext}
\usepackage[slantedGreek]{newtxmath}
\usepackage{multirow}
\usepackage{tablefootnote, footnote}
\usepackage{color, soul}

\graphicspath{{img-final/}}
\newcommand{\amp}{\&}

\author[Sam Morrell and Tim Naylor]{Sam Morrell\thanks{E-mail: \href{mailto:smorrell@astro.ex.ac.uk}{smorrell@astro.ex.ac.uk}} and Tim Naylor \\
School of Physics, University of Exeter, Exeter, EX4 4QL, UK
}

\date{Accepted 2019 August 03. Received 2019 August 03; in original form 2019 May 10}

\pubyear{2019}

\begin{document}
\label{firstpage}
\pagerange{\pageref{firstpage}--\pageref{lastpage}}
\title[The Radius of M-dwarfs]{Exploring the M-dwarf Luminosity - Temperature - Radius Relationships using Gaia DR2}
\pagerange{\pageref{firstpage}--\pageref{lastpage}}
\maketitle

\begin{abstract}
There is growing evidence that M-dwarf stars suffer radius inflation when compared to theoretical models, suggesting that models are missing some key physics required to completely describe stars at effective temperatures $(T_{\rm SED})$ less than about 4000K. 
The advent of Gaia DR2 distances finally makes available large datasets to determine the nature and extent of this effect. 
We employ an all-sky sample, comprising of $>$15\,000 stars, to determine empirical relationships between luminosity, temperature and radius. 
This is accomplished using only geometric distances and multiwave-band photometry, by utilising a modified spectral energy distribution fitting method. 
The radii we measure show an inflation of $3 - 7\%$ compared to models, but no more than a $1 - 2\%$ intrinsic spread in the inflated sequence. 
We show that we are currently able to determine M-dwarf radii to an accuracy of $2.4\%$ using our method.
However, we determine that this is limited by the precision of metallicity measurements, which contribute $1.7\%$ to the measured radius scatter. 
We also present evidence that stellar magnetism is currently unable to explain radius inflation in M-dwarfs. 
\end{abstract}


\section{Introduction} 
\label{sec:introduction}

It has been clear for a long time that something may be seriously wrong with our models of main-sequence M-dwarfs.
Historically it has been colour-magnitude diagrams (CMDs) of open clusters \citep[e.g.][]{Stauffer:2007aa,Bell:2012aa,Thompson:2014aa} which have demonstrated these problems, showing that the observed colours for stars below 0.7M${_\odot}$ can be up to half a magnitude redder than the model colours.
Problematically, CMDs give little insight as to where in the models the problems lie, because from a physical point of view colours and magnitudes are far removed from the fundamental parameters predicted by models.
The primary outputs from these models are the radius ($R$), effective temperature ($T_{\rm eff}$) and luminosity ($L$) for a star of a given mass. 
Hence measuring these quantities more directly is a promising way of testing the models, and in recent years such techniques have become sufficiently precise to be useful in this context.   

\subsection{Measuring the radii of M-dwarfs}
\label{sec:measuring}

Currently there are three commonly used methods to measure the radii of M-dwarfs, to which this paper seeks to add a fourth.
The double-lined eclipsing binaries (DEBs) provide a direct, fundamental test of the models because they yield radii at a given mass.
Again it has long been known that the eclipsing binaries show that M-star radii appear to be inflated for their mass \citep[see for example Figure 2 of][]{Chen:2014aa, Chaturvedi:2018aa,Parsons:2018aa,Mann:2019aa}, though \cite{Torres:2013aa} highlighted that the effect is subtle.
Furthermore there can be significant discrepancies between parameter determinations for the same binary, see for example the discussion of T-Cyg1-12664 in \cite{Han2017}, or the difference in measured radius for the secondary in PTFEB132.707+19.810/AD 3814 \citep{Kraus:2017aa, 2017ApJ...849...11G}. 
Nevertheless, at face value the failure of the models to match the eclipsing binaries in the mass-radius plane implies that the stellar structure models are incorrect.
However, there is a concern that binarism may affect stellar structure; specifically that if stars are tidally locked their high rotation rate will inflate them, and indeed some authors find evidence for this \citep[e.g. the careful work of][]{Kraus:2011aa}.
Hence the inflated radii measured for eclipsing binaries may not indicate a problem in single star models.

An alternative test for radius inflation is to obtain radii for the bright M-stars which are accessible to interferometers.
Given astrometric distances, interferometry provides a direct measure of the radius, but no direct measure of the mass.
Instead we must use one of the remaining parameters ($T_{\rm eff}$ or $L$) as a surrogate for mass, and it is probably $L$ which most closely follows mass and has little dependence on radius.
The luminosity is dictated by the thermodynamic properties in the core, and whilst this is dependent to a small degree on the radius, practically the central concentration of the star ensures that the dependence is weak.
Conversely the temperature is decided by the outer layers of the star and so has a strong dependence on radius.
Hence \cite{Boyajian:2012aa} measure luminosities (from broad-band photometry) and interferometric radii for a sample of 21 M-stars, and conclude that in the radius-luminosity plane the stars are inflated by about 5 percent, though again numbers are small.

The work of \cite{Mann:2013aa} and \cite{Mann:2015aa} overcomes the small-number-statistics problem by taking a sample of 183 M-stars with known distances and measuring their luminosities (again from broad-band photometry).
They use optical spectra to measure temperatures, which can then be converted into radii using the definition of effective temperature.
Their temperatures are derived by matching the lines and bands in their observed spectra to those of model atmospheres, hence we refer to this as a spectroscopic temperature, $T_{\rm sp}$.
This relies on the detailed opacities of individual lines being correct, as well as the structure of the model atmosphere.
Such temperature measurements can depend on the wavelength range and chemical species from which they are determined, which is especially problematic for M-dwarfs. 
For example determinations of temperature from optical \citep{Mann:2015aa, Cortes-Contreras:2017aa} and near infrared \cite{Newton:2014aa} spectra can differ by up to 1.5 spectral sub-types; around $5 - 10\%$ in $T_{\rm sp}$. 
At the root this technique relies on $T_{\rm eff}$ being equal to $T_{\rm sp}$.
That said, when plotted in the luminosity-temperature plane there is evidence either that the binaries are inflated with respect to the \citet{Mann:2015aa} sample (which is itself inflated with respect to the models) or there is a mismatch in the temperature scale. 

In summary there are three methods to relatively directly measure M-dwarf radii, each suffering to greater or lesser degree from problems of small-number statistics, rapid rotation or questions of how closely $T_{\rm sp}$ matches $T_{\rm eff}$.
Hence, in this paper we introduce a fourth method of measuring the radius, using the overall shape of the spectral energy distribution (SED) to measure what we will call $T_{\rm SED}$.
By fitting the broad-band photometry of a star with synthetic photometry of a model atmosphere we can derive both $T_{\rm SED}$ and, given the distances from Gaia DR2, the luminosity.
This allows us to create a sample of 15\,274 single stars which gives us the statistics to show that single stars and stars in binaries show a similar degree of inflation with respect to the models.
However, whether $T_{\rm SED}$ and $T_{\rm sp}$ are in agreement remains for debate. 

In addition this large sample allows us to address three further problems.

\subsection{Is there a correlation between [M/H] and radius?}

Metallicity is well known to have an effect on the structure of stars, and as we will discuss in Section \ref{sub:metallicity} theoretical models predict that there will be a 16 percent change in M-dwarf radius per dex change in [M/H] at a given luminosity.
The observational evidence for this dependency is weak, though it does seem to preclude a dependency much stronger than predicted, with the data in Figure 9 of \cite{Parsons:2018aa}  ruling out changes of more than about 10\% dex$^{-1}$ as a function of mass (which should correspond roughly to luminosity). \cite{Mann:2015aa} find a correlation of only about 4\% dex$^{-1}$ if the change in radius is measured as a function of absolute magnitude (which again should correspond roughly to luminosity).
The situation is very different if the radius change is measured as a function of effective temperature, with the same \cite{Mann:2015aa} data yielding 40\% dex$^{-1}$, which is similar to the result of \cite{Rabus:2019aa}, with an even larger dependence being found for the lowest metallicity M-dwarfs by \cite{Kesseli2019}. 
In fact these results simply show that from a physical point of view measuring the radius change as a function of temperature confuses the issue, since changing the radius at a fixed luminosity will of itself change the temperature.
In this work when addressing metallicity we will, therefore, measure the radius change as a function of luminosity, and using this will show in Section \ref{sub:metallicity}, that our technique shows no relationship between radius and metallicity, but our upper limit is consistent with the theoretical predictions.

\subsection{What causes the radius inflation?}
\label{sub:intro-causes-of-radius-inflation}

Given that convective flows are the dominant means of energy transport out of the star, many have made the reasonable assertion that inhibiting convection would cause an increase in the stellar radius, and hence explain the inflation of M-dwarfs. 
This inhibition could be accomplished by the strong magnetic fields \citep{Mullan:2001aa,MacDonald:2017ab} that are 
thought to be the fundamental drivers of both magneto-convection and the delay of convective onset \citep[e.g.][]{MacDonald:2014aa}.
Hence there has been much theoretical work over the last decade to develop consistent stellar evolution models that correctly account for dynamo effects and stellar magnetism, including the resulting starspots \citep{Mullan:2001aa,Feiden:2013aa,Somers:2016aa,MacDonald:2017ab}. 
Recently these effects have been modelled in 1D stellar structure models by \citet{Ireland:2018aa}, who adopt a depth dependent mixing length theory parameter $\alpha_{\rm MLT}$; emulating the effect of convective inhibition. 
In their work they observe radii inflated by 10 - 15\% when compared to models that do not treat convection in this way, though they caution that such treatments of magnetic inhibition are highly uncertain and may be difficult to calibrate. 

Despite this success, the hypothesis that magnetic fields cause inflation has been brought into question by \citet{Kesseli:2018aa}, who used a $v$sin($i$) technique to show that the radii of samples of rapidly-rotating stars are consistent (to within $5\%$) with those of slowly-rotating stars, and the inflation in eclipsing binaries consistent with single stars. 
In addition \citet{Kochukhov:2019aa} conclude that magnetic inflation models do not support their observations of the DEB YY Gem.

Theories which invoke magnetism to explain the radius inflation of M-dwarfs make two predictions we will test in this paper.
The first is that unless M-dwarfs all have very similar fields, at a given mass (or luminosity) they should have radii which range between the predictions of the non-magnetic models and some maximum inflation.
Our sample is sufficiently large to show that any spread in M-dwarf radii is much less than $1 - 2\%$ (\autoref{sub:discussion-radius-inflation}), apparently ruling out the magnetic inflation models.
The second prediction is that the degree of radius inflation should correlate with magnetic activity indicators \citep{Lopez-Morales:2007aa}.
Again we find no such relationships (\autoref{sec:activity}), but for indicators such as X-ray activity this could be explained if the surface field is unrelated to the magnetic field in the bulk of the star \citep[see][for a discussion of how under-developed models of M-stars are]{2017LRSP...14....4B}.  However this criticism cannot be levelled at the absence of a correlation between rotation period (or Rossby number) and inflation we find, which again suggests magnetic fields may not be responsible for radius inflation.

\subsection{Radii of planets}
Whilst understanding the structure of M-stars is a significant problem in its own right, it is currently being given added impetus by the need for precise predictions of M-dwarf radii to understand the structure of rocky exoplanets.
Their small size makes M-stars the current target-of-choice for rocky exoplanet hunting, since it maximises both the radial velocity and transit signal from a planet of a given mass.
However, to use the transit depth to obtain planetary radii and hence densities requires precise M-star masses.
The technique developed in this paper gives just such precise radii which can be obtained from archive data, though it can be improved by the addition of a spectroscopic metallicity.


\section{Method}
\label{sec:method}
Given a precise parallax one can integrate the area beneath the SED to find the luminosity of the star, while the shape of the SED is a function of temperature. 
With these, the radius of the star can be calculated. 
Importantly, both are a function of only the stellar photosphere; allowing the fitting process to be uncoupled from the model interiors. 
Hence using just synthetic photometry from model atmospheres we have developed a method that uses broadband photometry, readily available from public surveys, to sample the SED and infer these properties. 
\input{tab/data-sources.tex}
The bands that we used, which comprised broadband photometry from optical, near-infrared and mid-infrared surveys, are detailed in \autoref{tab:data-sources}.
\begin{figure}
	\includegraphics[width=\columnwidth]{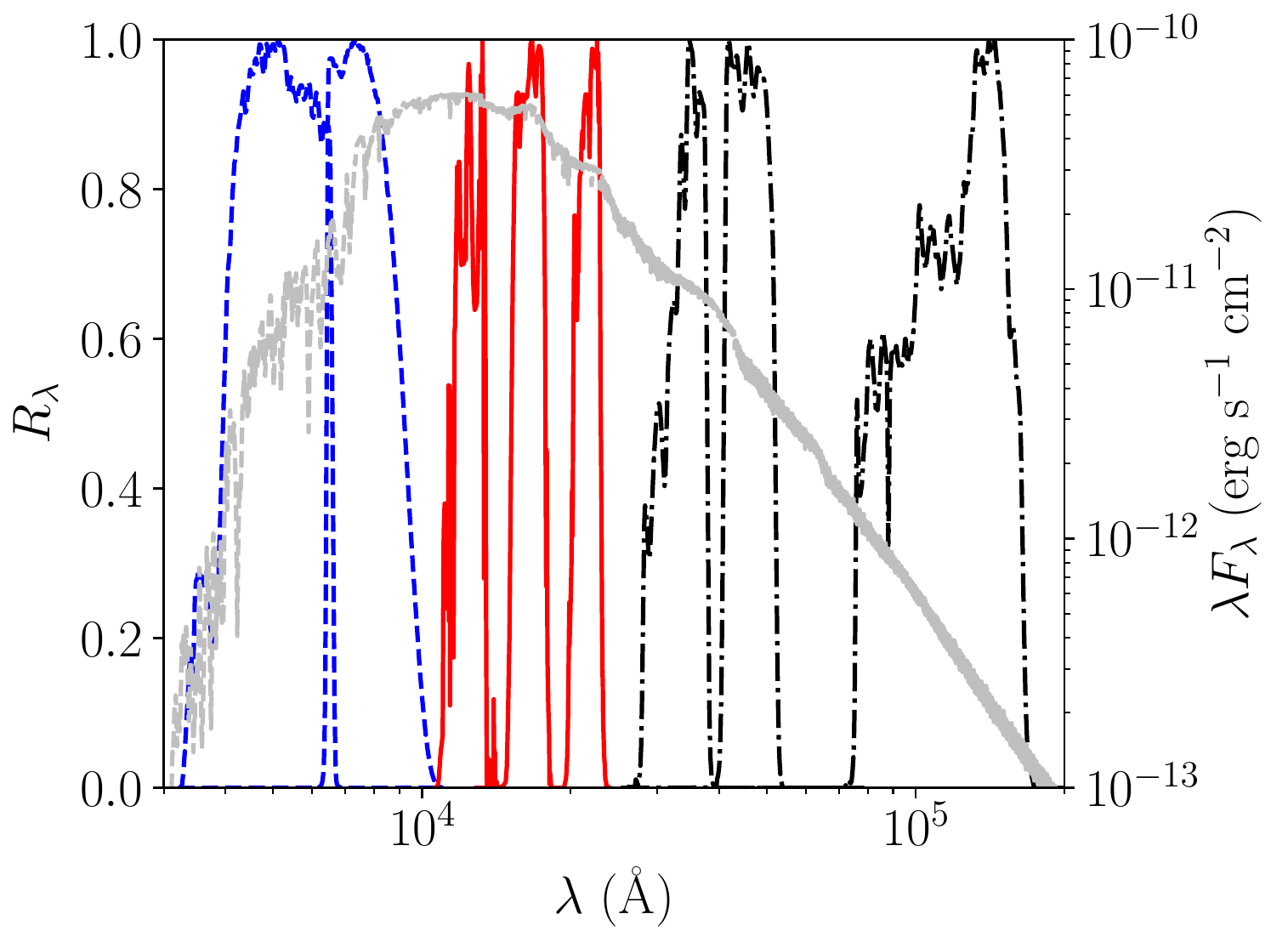}
	\caption{The system responses used to generate the synthetic photometry. The photometry is comprised of magnitudes from Gaia (blue dashed), the Two Micron All-Sky Survey (red solid) and WISE (black dot-dashed). For reference, the model spectrum for an M-dwarf star with an effective temperature $T_\text{eff} = 3300$K is included (grey dashed). }
	\label{fig:filter-coverage}
\end{figure}
The system responses used to generate our synthetic photometry are plotted in \autoref{fig:filter-coverage}, along with a model spectrum of an M-dwarf star.
This helps justify our motivation for the choice of data and photometric systems. 
First, all photometric systems correspond to all sky surveys whose system responses are well understood. 
This means that not only can we draw a sample of stars from the entire sky, but the characterisation of the system responses means that our folded photometry will closely replicate the original photometry. 
Using these surveys, we also have excellent coverage of the stellar SED, with the optical Gaia prism photometry \citep{Gaia2016,Gaia2018} sampling blue-ward of the blackbody peak, 2MASS NIR photometry \citep{2mass} sampling the peak itself and AllWISE photometry \citep{Wright:2010aa} constraining the Rayleigh-Jeans tail down to fluxes 3 orders of magnitude lower than the peak.
This combination means that we have the ability to accurately constrain the colour of the objects, and thus the $T_{\rm SED}$, and make a robust estimate of the luminosity of a star from its entire SED. 

\subsection{Target Selection} 
\label{sub:target_selection}
Each catalogue used is shown in \autoref{tab:data-sources} along with the relevant publications. 
We began by selecting all stars from Gaia DR2 with $r_{\rm est} < 100$pc from \citet{Bailer-Jones:2018aa}. 
We chose this distance cut because the systematic uncertainties in the Gaia DR2 parallaxes generally remain below $0.1$mas \citep{Lindegren:2018aa}.
This constraint guarantees that our distances are not affected by Gaia's astrometric systematics while remaining good to $1\%$ uncertainty in luminosity. 
This preliminary sample comprised 138\,279 sources. 
Each photometric catalogue was then cross matched with this preliminary sample and the quality cuts shown in \autoref{tab:data-sources} applied. 
We omitted the $W_4$ band from the fitting process due to its poor signal-to-noise, and because $W_3$ does an adequate job of characterising the Rayleigh-Jeans tail of even the coolest stars in the sample. 
The final input catalogue is then constructed by combining the photometry for only the stars common to all source catalogues. 
The number of sources remaining in the final input catalogue are 15\,765. 

\subsection{Model Grid} 
\label{sub:model_grid}
The synthetic photometry was produced using the BT-Settl CIFIST stellar atmosphere grid\footnote{\href{https://phoenix.ens-lyon.fr/Grids/BT-Settl/CIFIST2011_2015/}{https://phoenix.ens-lyon.fr/Grids/BT\-Settl/CIFIST2011\_2015/}} \citep{Allard:2012aa}. 
These atmospheres are provided in units of mean disc intensity at the stellar surface $I_\lambda$, which includes the effect of limb darkening.
$I_\lambda$ is related to the flux observed at the surface of the Earth $F_\lambda$ by 
\begin{equation}
	F_\lambda = I_\lambda \frac{R^2}{d^2},
\end{equation}
where $R$ is the stellar radius, which we fitted, and $d$ is its distance from the Earth \citep[e. g.][]{Girardi:2002aa}. 
Strictly, this equation should also have a factor that would account for extinction $A_\lambda$. 
However, as our entire sample is within $100$pc, the effect of extinction on our photometry is negligible. 
To reduce computational time we downsampled the wavelength scale of the input stellar atmospheres to $\simeq$1\AA\ bins using a flux-conserving algorithm.
We then performed a bilinear interpolation, first in $\log(g)$ then in $T_{\rm eff}$, on the 4 nearest downsampled models to produce the final $I_\lambda$. 
We folded $I_\lambda$ through the system responses to yield the mean apparent synthetic magnitude of a unit surface area of the model atmosphere at its surface within the $i^{\rm th}$ band $Z_i$ with 
\begin{align}
	m_{i, \text{syn}} &= - 2.5 \log_{10} \left[\frac{\int_\lambda F_\lambda S_{\lambda, i} d\lambda}{\int_\lambda f^\circ_{\lambda, i} S_{\lambda, i} d\lambda}\right] + m^\circ_i \nonumber \\
	& = - 2.5 \log_{10} \left[ \frac{\int_\lambda I_\lambda S_{\lambda, i} d\lambda}{\int_\lambda f^\circ_{\lambda, i} S_{\lambda, i} d\lambda}\right] - 5\log_{10} \left[ \frac{R}{d}\right] + m^\circ_i \nonumber \\
	& = Z_i - 5 \log_{10} \left[ \frac{R}{d} \right], 
\end{align}
where $S_{\lambda, i}$, $f^\circ_{\lambda, i}$ and $m^\circ_i$ are the system response, zero point flux and zero point of the $i^{\rm th}$ band. 

\subsection{Free Temperature-Radius Fitting} 
\label{sub:photometric_fitting}
For the fitting, we adapted the spectral energy distribution fit (SEDF) method \citep[e.g. ][]{Pecaut:2013aa,Masana:2006aa} with an addition to remove the dependence on \textit{a priori} knowledge of the angular radius of the star $\theta$ (see \autoref{sub:analytical-determination-of-r}).
This method is similar in principle to the infrared flux method of \citep{Blackwell:1977aa, Blackwell:1979aa}, reviewed in \citet{Casagrande:2008ab}. 
However it is applied across the entire SED, thus generalises well to a wide range of spectral types; including cool stars like M-dwarfs. 
Essentially we used the shape of the SED to measure the temperature.  

We compared our synthetic photometry $m_{i, \text{syn}}$ to the photometric data $m_i$ using 
\begin{equation}
	\chi^2 = \sum^N_i \left( \frac{m_i - m_{i, \text{syn}}}{\sigma_i} \right)^2
	= \sum^N_i \left( \frac{m_i - Z_i + 5\log_{10}(R / d)}{\sigma_i} \right)^2, 
	\label{eq:rt-chisq}
\end{equation}
where $\sigma_i$ is the uncertainty in the $i{\rm th}$ photometric band and $i = G_\text{BP}$, $G_\text{RP}$, $J$, $H$, $K_\text{s}$, $W1$, $W2$ and $W3$. 
For $\sigma_i$ we adopted a floor value of 0.01 mag, which corresponds to roughly $1\%$, for all photometric uncertainties in the entire sample. 
The free parameters for the fit are $R/d$, $T_{\rm eff}$ and $\log(g)$, which we explored using a simple grid search to generate a 3D cube of $\chi^2$. 
The minimum value of $\chi^2$ within the grid determines the best fitting solution. 
We constrained the $\log(g)$ axis by applying a tophat prior of $\pm 0.5$ dex around the prescribed model $\log(g)$, which was found by matching $M_G$ of each star with the \citet{Baraffe:2015aa} 4Gyr isochrone. 

\begin{figure}
	\includegraphics[width=\columnwidth]{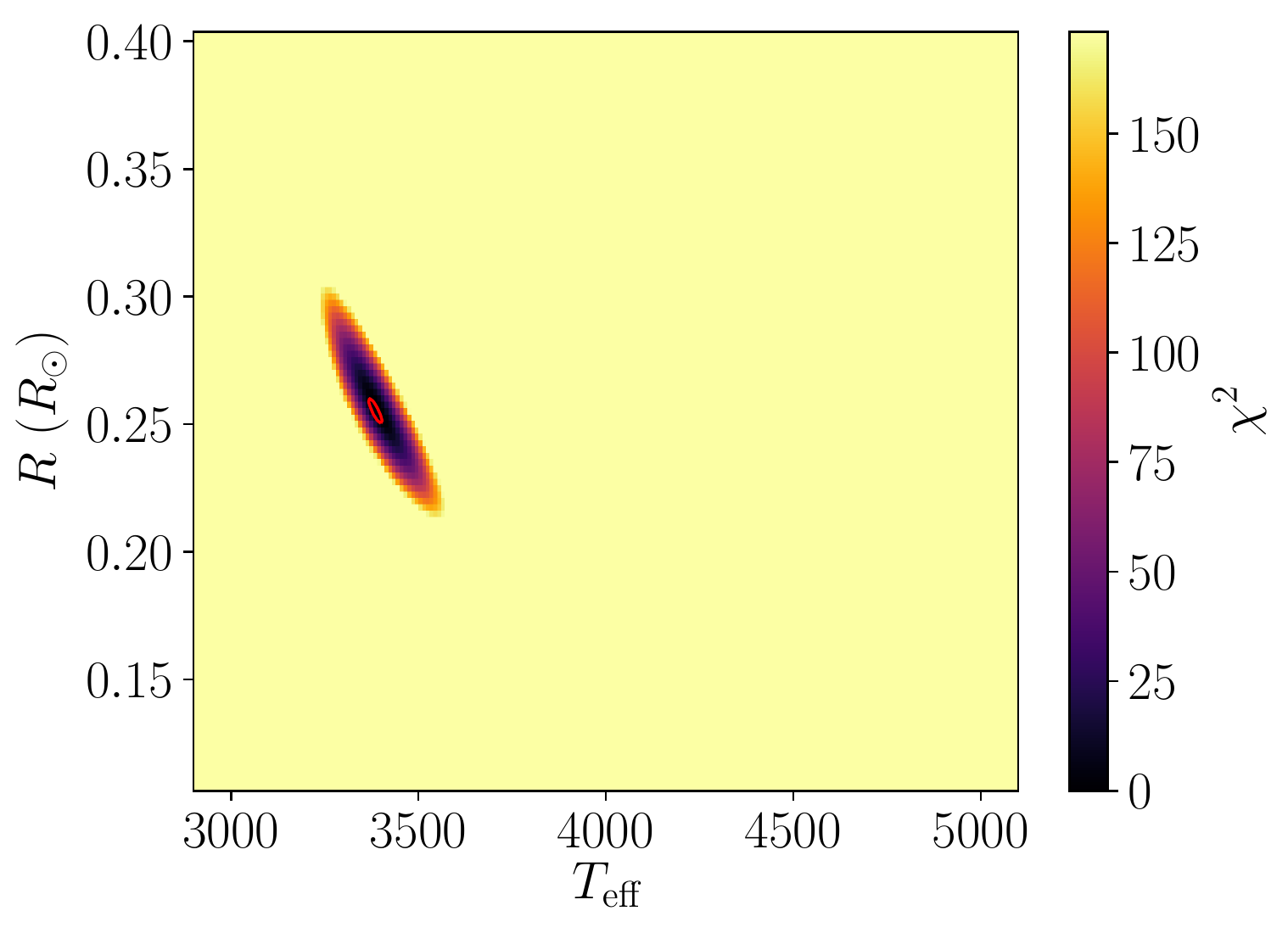}
	\caption{The search space for one of our targets whose $\chi^2$ lies at the median value of the randomly selected uncertainty sample. The red ellipsoid indicates the 68\% confidence contour resulting from the process. }
	\label{fig:space-splot}
\end{figure}

The grids were first transformed into probability space, using $P(T_{\rm SED}, \log(g), R/d) = \exp{( - \chi^2 / 2)}$, and normalised. 
The 2D $\chi^2$ space necessary for producing confidence contours was then calculated by marginalising the 3D cube over $\log(g)$. 
We next converted $R/d$ to $R$ using the geometric distances of \citet{Bailer-Jones:2018aa}, and allowed for uncertainty in distance by convolving each row of constant $T_{\rm eff}$ in the 2D $\chi^2$ space with a Gaussian whose standard deviation was the mean uncertainty from the lower and upper distance bounds; as these are nearly symmetrical. 
The confidence contours were determined from the resulting 2D PDF by identifying the set of highest probability pixels whose sum was 0.68 and drawing a contour around them. 
We obtained the uncertainties in $T_{\rm SED}$ and $R$ from a 1D distribution of probability by marginalising along the remaining axis. 
To illustrate the correlation between $R$ and $T_{\rm eff}$, we include an example plot of the $\chi^2$ space resulting from a fit to one of our targets in \autoref{fig:space-splot}. 
We randomly picked 158 stars from our input catalogue, 1\% of our sample, upon which we performed this full grid search.
We used this sample to determine uncertainties characteristic of the full sample.
The 68\% confidence contours from which those uncertainties are derived are shown in  \autoref{fig:uncer-contours}.

\subsection{Analytical Determination of \texorpdfstring{$R$}{Radius}} 
\label{sub:analytical-determination-of-r}
\begin{figure*}
    \centering
    \includegraphics[width=.32\textwidth]{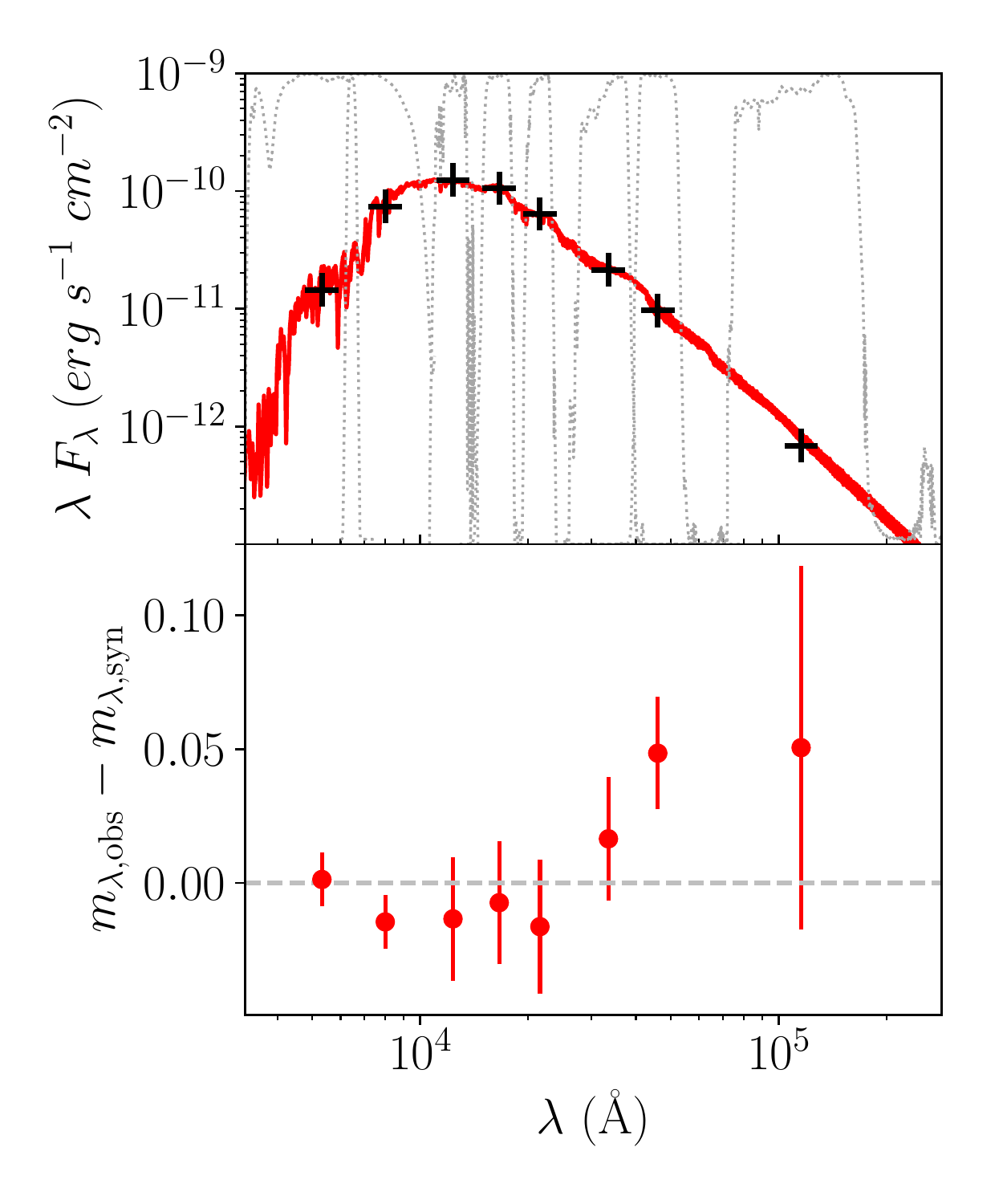}\hfill
    \includegraphics[width=.32\textwidth]{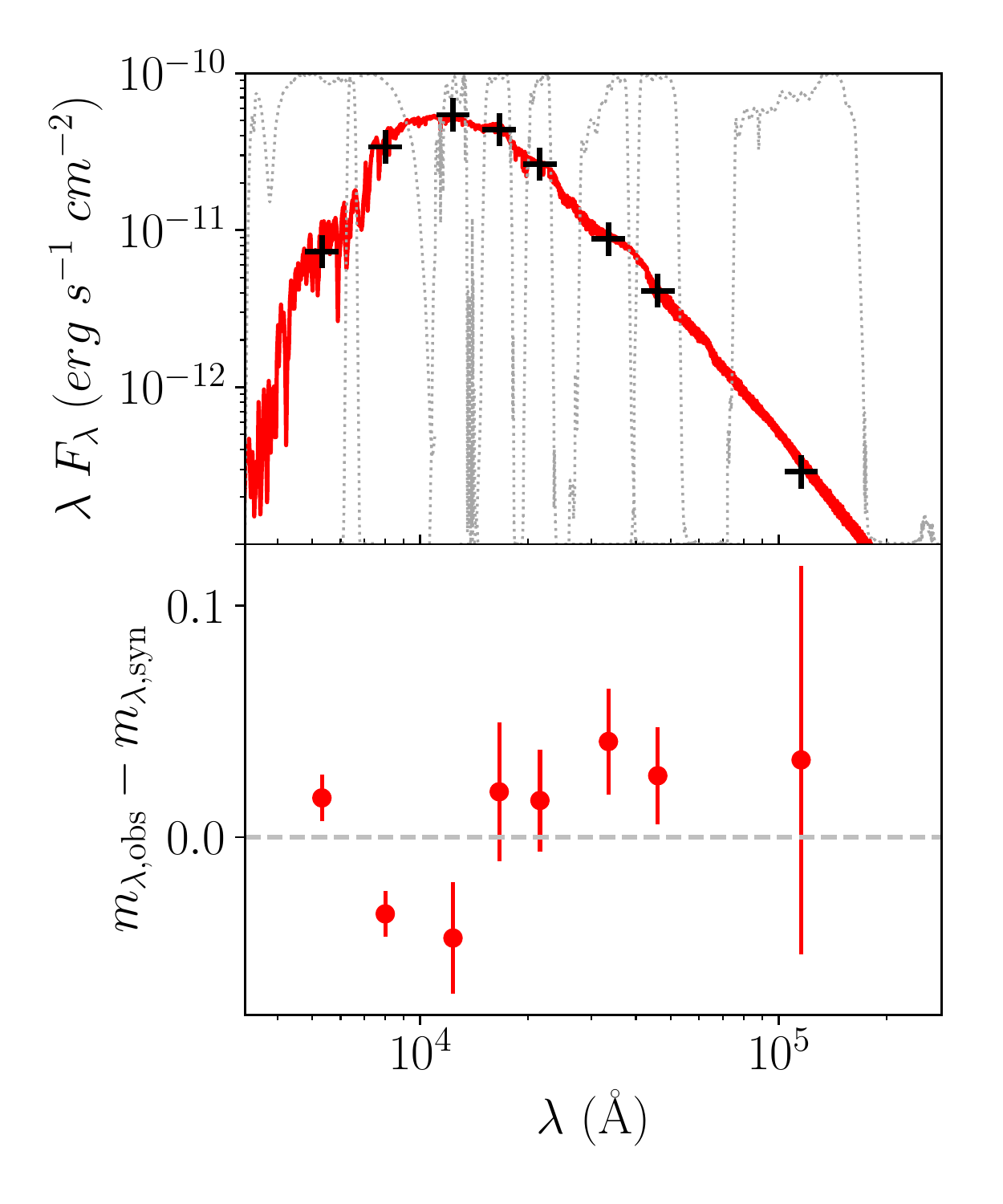}\hfill
    \includegraphics[width=.32\textwidth]{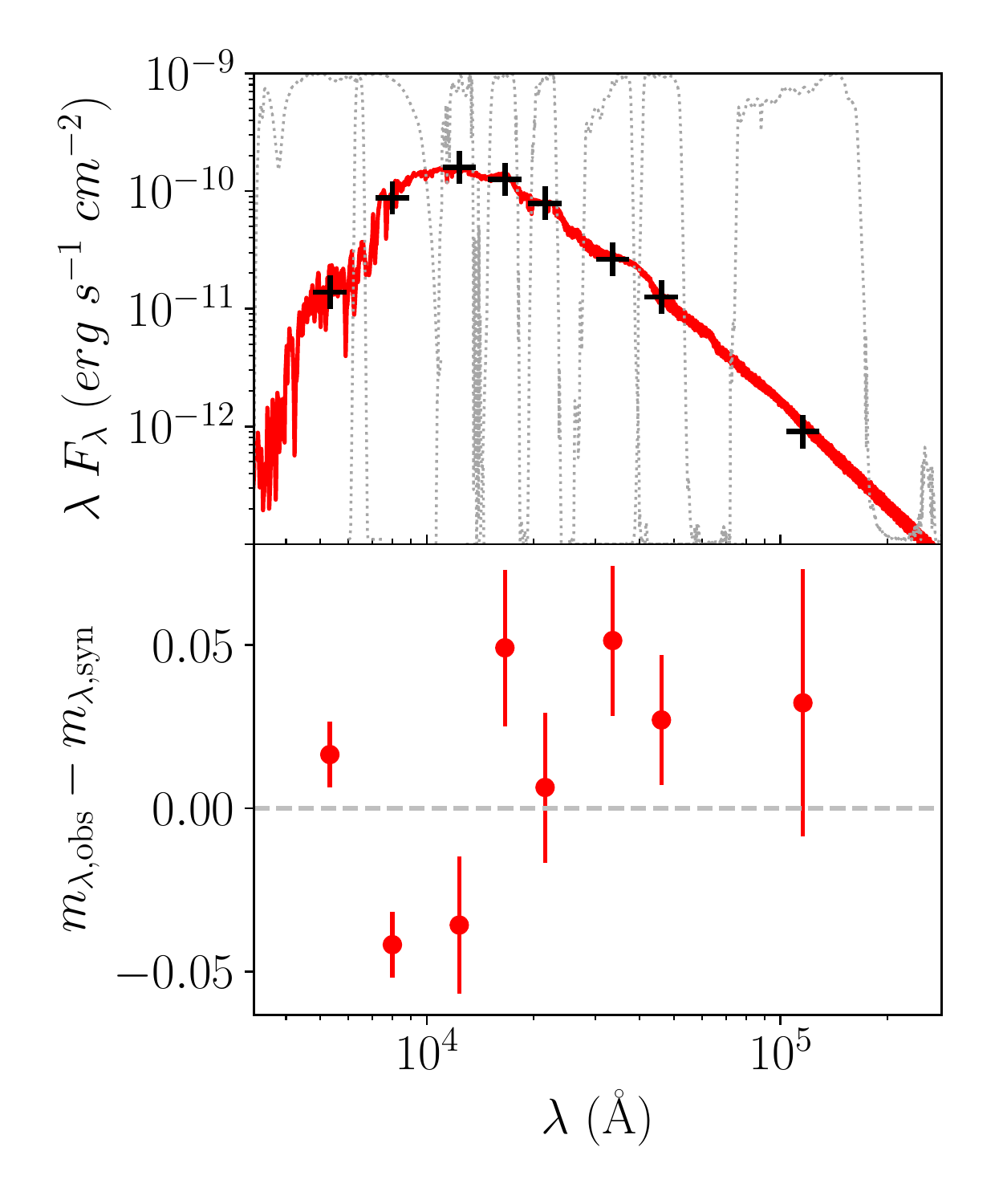}
    \caption{Fits resulting from the use of the method presented in \autoref{sub:photometric_fitting} and \autoref{sub:analytical-determination-of-r}. The best fitting model spectrum for each target is shown in the top panel (red) with the observed photometry from which it was derived overlaid (black). The appropriate bandpasses are plotted in light grey for reference. In the bottom panel are the residuals and uncertainties in magnitudes for each photometric band. The middle panel corresponds to the median $\chi^2$ of our randomly selected uncertainty sample, while the left and right panels show the stars at the lower and upper $68\%$ density bounds respectively. }
    \label{fig:lam-flam-plots}
\end{figure*}

Performing a full 3D grid search on our entire input catalogue is intractable due to computational time constraints.
So for a particular $T_{\rm eff}$ and $\log(g)$ we determined best fitting $R$ by analytically minimising $\chi^2$, effectively making our search space 2D. 
We accomplished this by differentiating \autoref{eq:rt-chisq} with respect to the $5\log_{10}(R / d)\ $ term and finding the stationary point of the derivative to analytically minimise $\chi^2$, yielding 
\begin{equation}
	\log_{10}\left(\frac{R^2}{d^2}\right) = \log_{10}\left(\theta^2\right) = - 0.4 \left(\sum^N_i \frac{ Z_i - m_i}{\sigma_i^2}\right) \bigg{/} \left(\sum^N_i \frac{1}{\sigma_i^2}\right). 
	\label{eq:rasr_on_dsqr_from_fit}
\end{equation}
This provides the dilution factor $R^2 / d^2$ and hence, by applying the distances from \citet{Bailer-Jones:2018aa}, the stellar radius $R$.
For each $\log(g)$ and $T_{\rm eff}$ we analytically minimised $\chi^2$ to find the radius in this way. 
This results in a 2D $\log(g) - T_{\rm eff}$ space. 
An example of the resulting $\chi^2$ space is shown in \autoref{fig:space-splot}, and example fits shown in \autoref{fig:lam-flam-plots}. 

\subsection{Flagging} 
\label{sub:flagging}
\begin{figure}
	\includegraphics[width=\columnwidth]{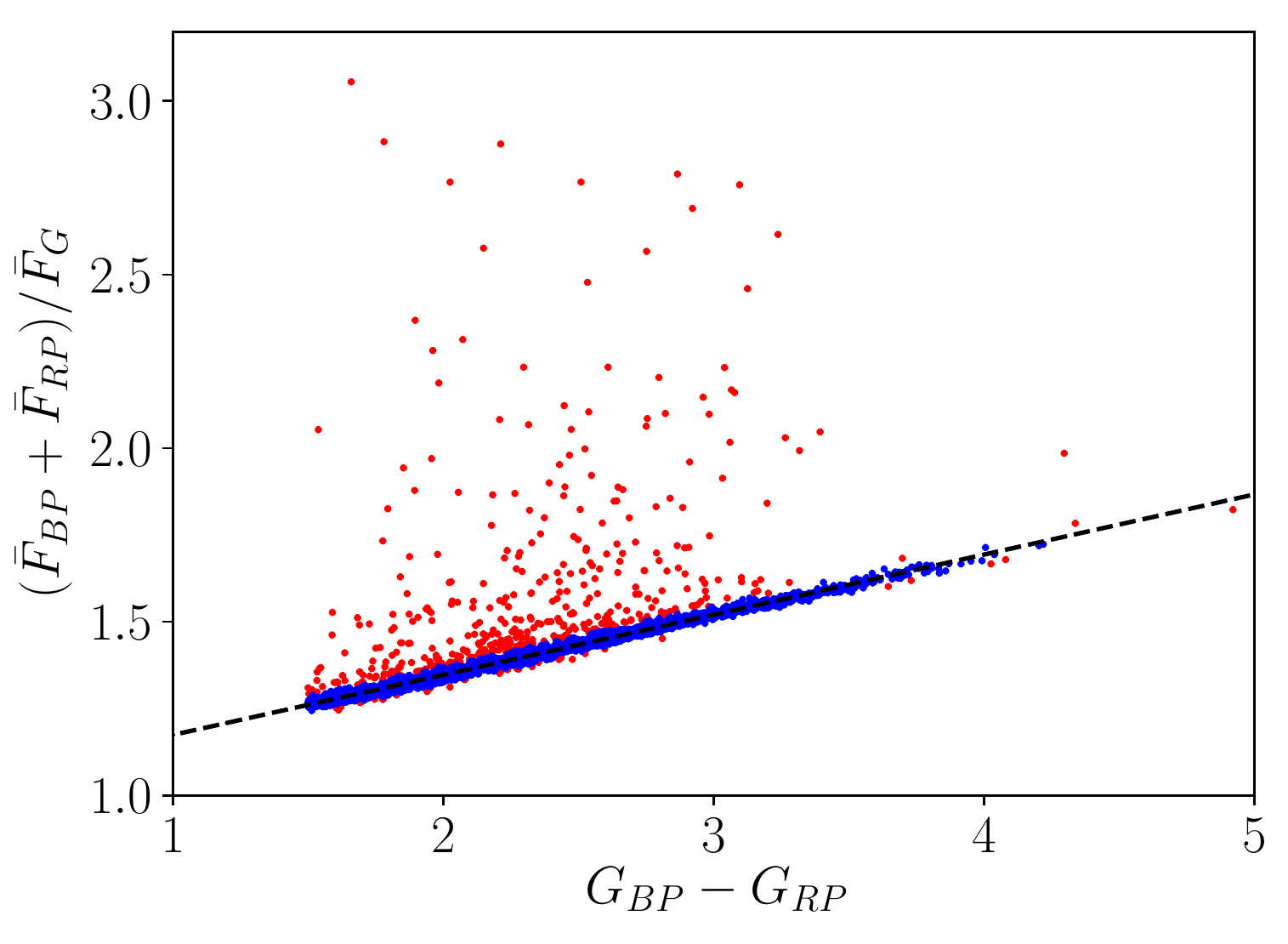}
	\caption{This figure illustrates the sigma clipping performed on our final catalogue. The black dashed line shows the final linear fit to the clipped sample after 8 iterations. The red points in this plot are those lying more than $5\sigma$ away from this line, and are thus flagged as \textbf{bad\_phot}. Those points remaining in blue lie within $5\sigma$ of the fit, and are considered to have uncontaminated Gaia photometry. }
	\label{fig:sigma-clip-phot-cut}
\end{figure}

In addition to the initial cuts we performed on the source catalogues, which already produced a stringently constrained sample, we also performed additional post processing to produce flags to be included with the fitted parameters. 
As the $G_{\rm BP}$ and $G_{\rm RP}$ fluxes are integrated over a $3.5 \times 2.5\ \mathrm{arcsec}^2$ field they are susceptible to contamination from both bright, nearby sources and sky background.  
Following the method of \citet{Evans:2018aa}, we applied the \textbf{bad\_phot} photometric contamination flag by $\sigma$-clipping sources in the flux excess ratio vs. colour plane. 
See \autoref{fig:sigma-clip-phot-cut} for the resulting classification after 8 iterations. 

The Gaia astrometric data can similarly suffer contamination from crowding and binaries, causing both exaggerated and errant parallaxes \citep{Lindegren:2018aa}.
For a similar sample in \citet{Lindegren:2018aa}, the criterion
\begin{equation}
	u < 1.2 \times \max( 1, \exp(-0.2 (G_G - 19.5)) )
	\label{eq:lindegren_bas_astr}
\end{equation}
was used to clean the sample of poor astrometry. 
Thus, any source that does not satisfy this expression is flagged with the \textbf{bad\_astr} flag in our final catalogue. 

There are also those sources for which the fitting could not converge on a reasonable solution, probably because their true $T_{\rm SED}$ lies outside the bounds of our sample space.
Thus sources lying on either $T_{\rm eff}$ bound are flagged as \textbf{bad\_teff} in the final catalogue.
Finally, those sources which remained unflagged were assigned the \textbf{good} flag, meaning their input data should be free of both photometric and astrometric contamination, and they should have a well constrained $T_{\rm SED}$. 

The full table, including the input data and resulting parameters, are available at CDS via anonymous ftp to cdsarc.u-strasbg.fr (130.79.128.5) or via \href{http://cdsarc.u-strasbg.fr/viz-bin/qcat?VI/156}{http://cdsarc.u-strasbg.fr/viz-bin/qcat?VI/156}, the University of Exeter ORE repository (\href{https://doi.org/10.24378/exe.1683}{https://doi.org/10.24378/exe.1683}) and the GitHub repository of supplementary material for this publication\footnote{\href{https://github.com/sammorrell/mn19-supplementary-material}{https://github.com/sammorrell/mn19-supplementary-material}}. 
The table columns are described in \autoref{tab:column-descriptions}. 
\input{tab/column-descriptions}

\section{Results} 
\label{sec:results}
\begin{figure*}
	\includegraphics[width=\textwidth, trim=0 0.5cm 0 0]{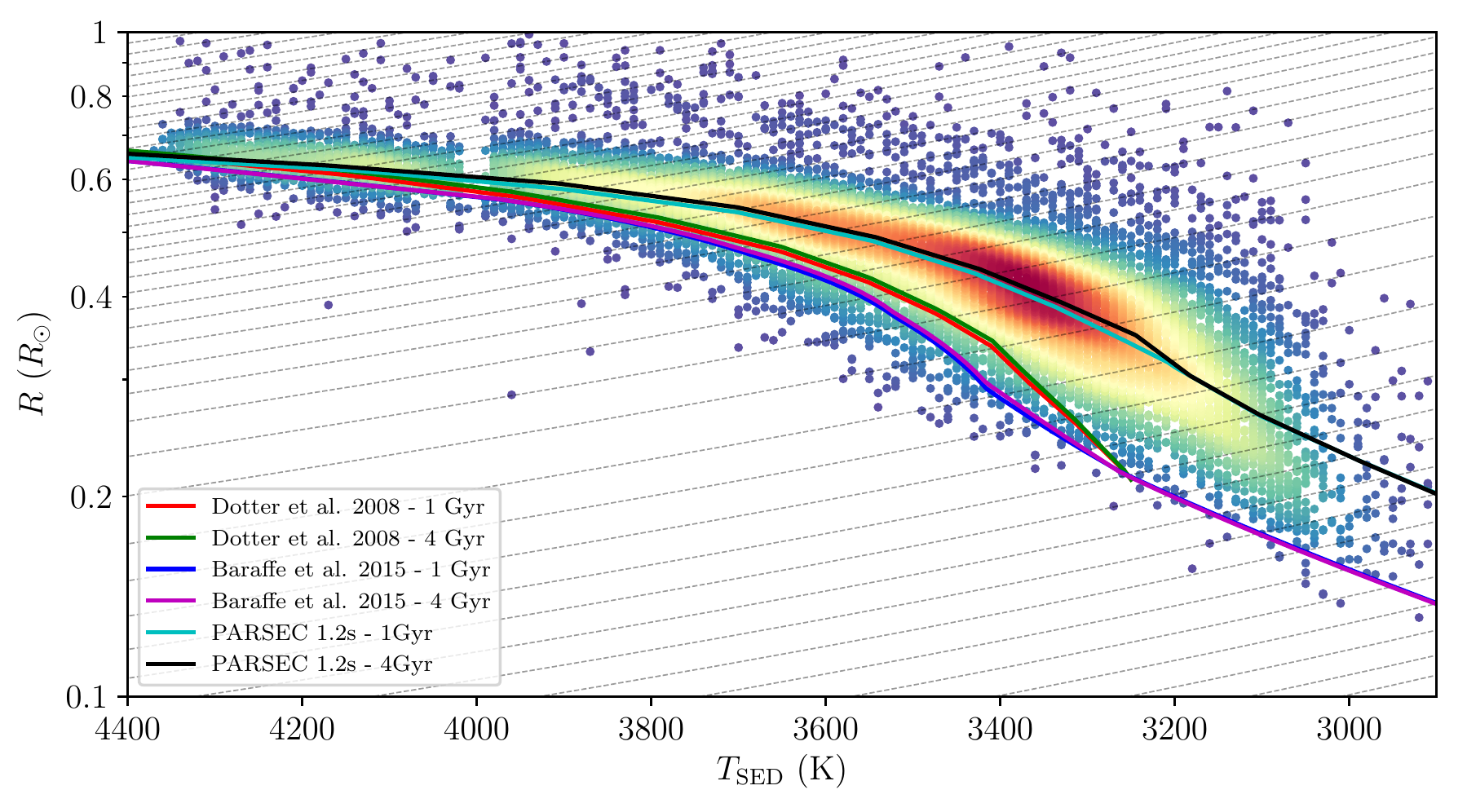}
	\caption{The distribution of points from our full sample of 15\,274 sources in $T_\mathrm{SED} - R$ space. The colour map results  from a kernel density estimation, which is intended to indicate the density of points within the plot. For comparison, we include the \citet{Dotter:2008aa} isochrones for z=0.018 at 1 and 4 Gyr, \citet{Baraffe:2015aa} isochrones for 1 and 4 Gyr and PARSEC 1.2S \citep{Chen:2014aa,Marigo:2017aa} isochrones for 1 and 4 Gyr in red, green, blue, purple, cyan and black lines respectively. Several points are clipped off the sides of the figure. The dashed lines trace constant luminosity, drawn every $100$K along the \citet{Baraffe:2015aa} isochrone. }
	\label{fig:rad-teff-const-lum}
\end{figure*}

\begin{figure}
	\centering
	\includegraphics[width=\columnwidth]{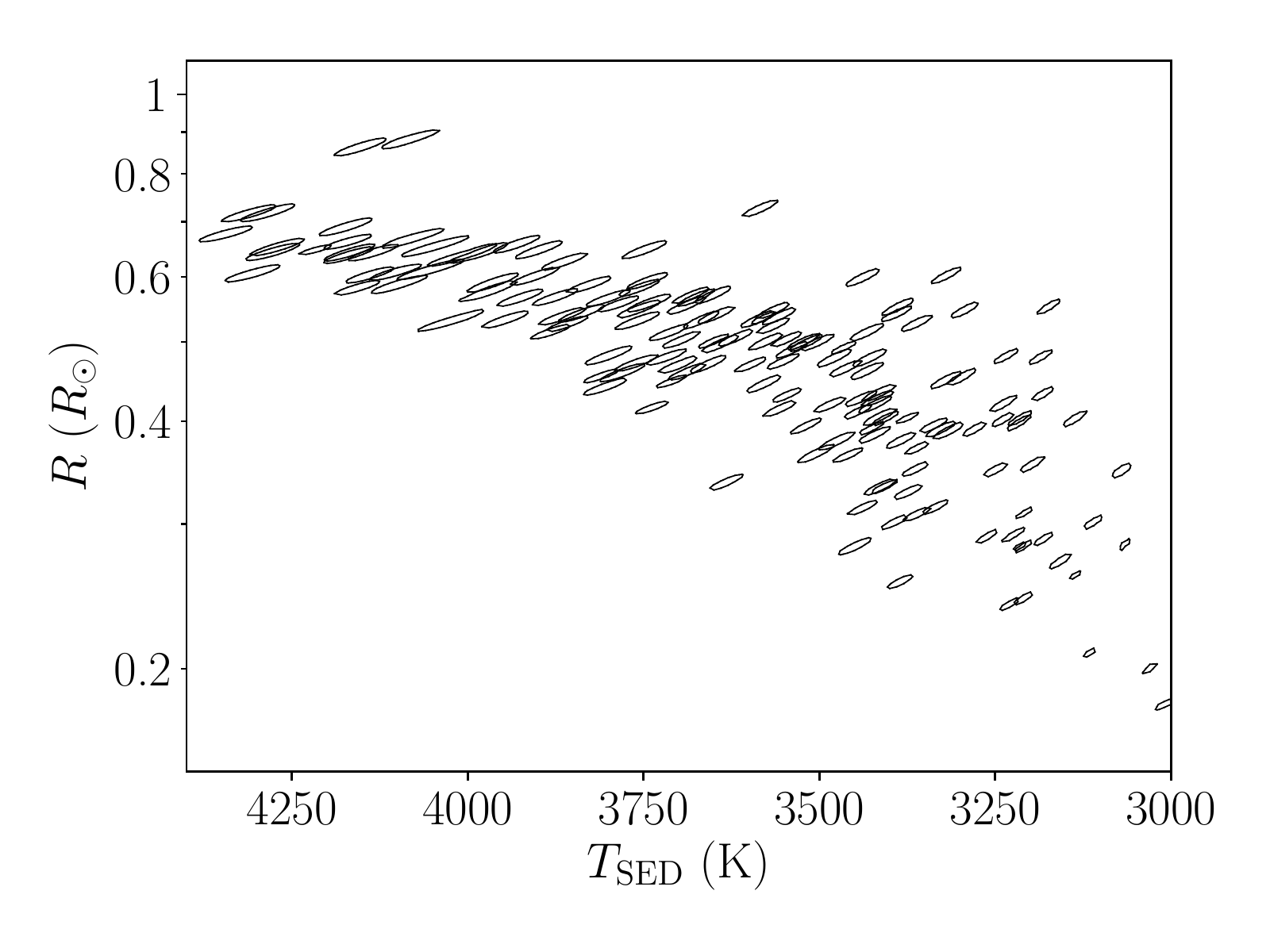}
	\caption{The 68\% confidence contours resulting from the full 3D grid search we performed on 1\% of our sample (158 stars) in \autoref{sub:photometric_fitting}.  }
	\label{fig:uncer-contours}
\end{figure}

Of the 15\,765 targets in the input catalogue, 15\,274 of them are flagged as \textbf{good}. 
We have plotted our sample in $R - T_\mathrm{SED}$ space in \autoref{fig:rad-teff-const-lum}, along with a selection of solar metallicity isochrones at 1 Gyr and 4 Gyr. 
The 68\% confidence contours in the $R-T_{\rm SED}$ space from the subset described in \autoref{sub:photometric_fitting} are shown in \autoref{fig:uncer-contours}. 
These show that we are able to determine the radius to a median statistical uncertainty of $0.009\ R_\odot$ $(1.6\%)$, ranging to a maximum uncertainty of $0.025\ R_\odot$ $(2.6\%)$.
We found the mean uncertainty in temperature $T_{\rm SED}$ was $35$K $(1.0\%)$, ranging to a maximum of $100$K $(2.7\%)$. 

\begin{figure}
	\includegraphics[width=\columnwidth]{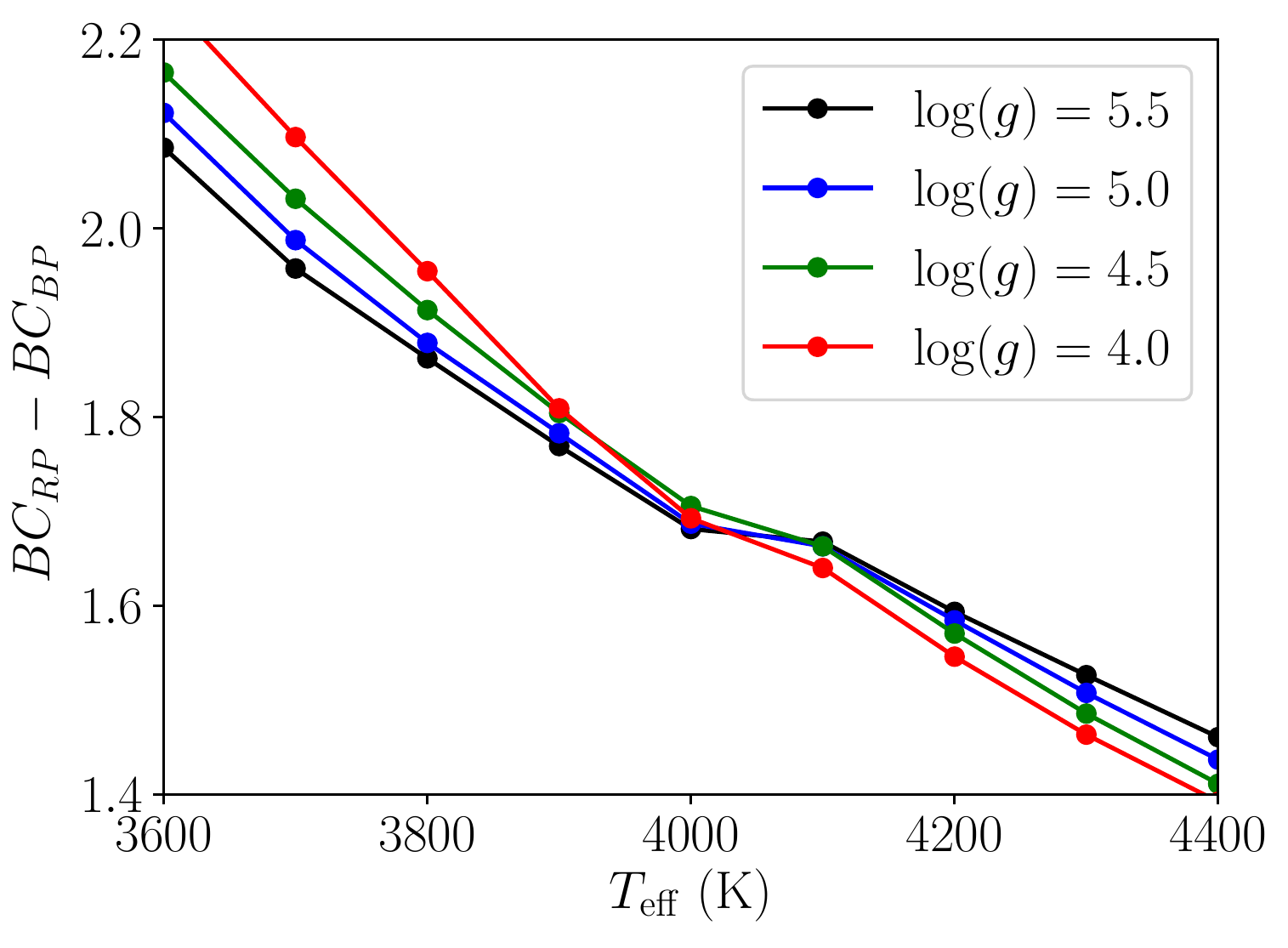}
	\caption{The gap in the stellar sequence at 4000K evident in \autoref{fig:rad-teff-const-lum} is due to a discontinuity in the CIFIST BT-Settl model grid. By plotting the bolometric corrections from the Gaia DR2 bands this effect is clearly seen. The plot shows $\log(g) = 4.0, 4.5, 5.0\ \mathrm{ and }\ 5.5$ (black, blue, green and red respectively) to demonstrate that this discontinuity affects the entire span of $\log(g)$ sampled by our grid. }
	\label{fig:model-discon-4000k}
\end{figure}

The gap in the stellar sequence at $4000$K in \autoref{fig:rad-teff-const-lum} is caused by a discontinuity in the CIFIST BT-Settl model grid where the monotonic relationship between bolometric correction, defined as $M_{\rm bol} - M_i$ where $M_i$ is the absolute magnitude of the $i^{\rm th}$ photometric band, and $T_{\rm eff}$ breaks down (\autoref{fig:model-discon-4000k}). 
We found that comparisons between observations and the SED resulting from the atmosphere at $4000$K produce a higher $\chi^2$ than those from the neighbouring atmospheres, causing our fitting to favour the SEDs produced from the atmospheres adjacent to that at $4000$K. 
That this is a property of the model, as opposed to the fitting process, is supported by the fact that when we perform fits with grids derived from different atmospheres, as with the sub and super-solar metallicity grids in \autoref{sub:metallicity}, the discontinuity disappears. 

\subsection{Radius Inflation}
\label{sec:radius-inflation}
\begin{figure*}
    \centering
    \includegraphics[width=\textwidth]{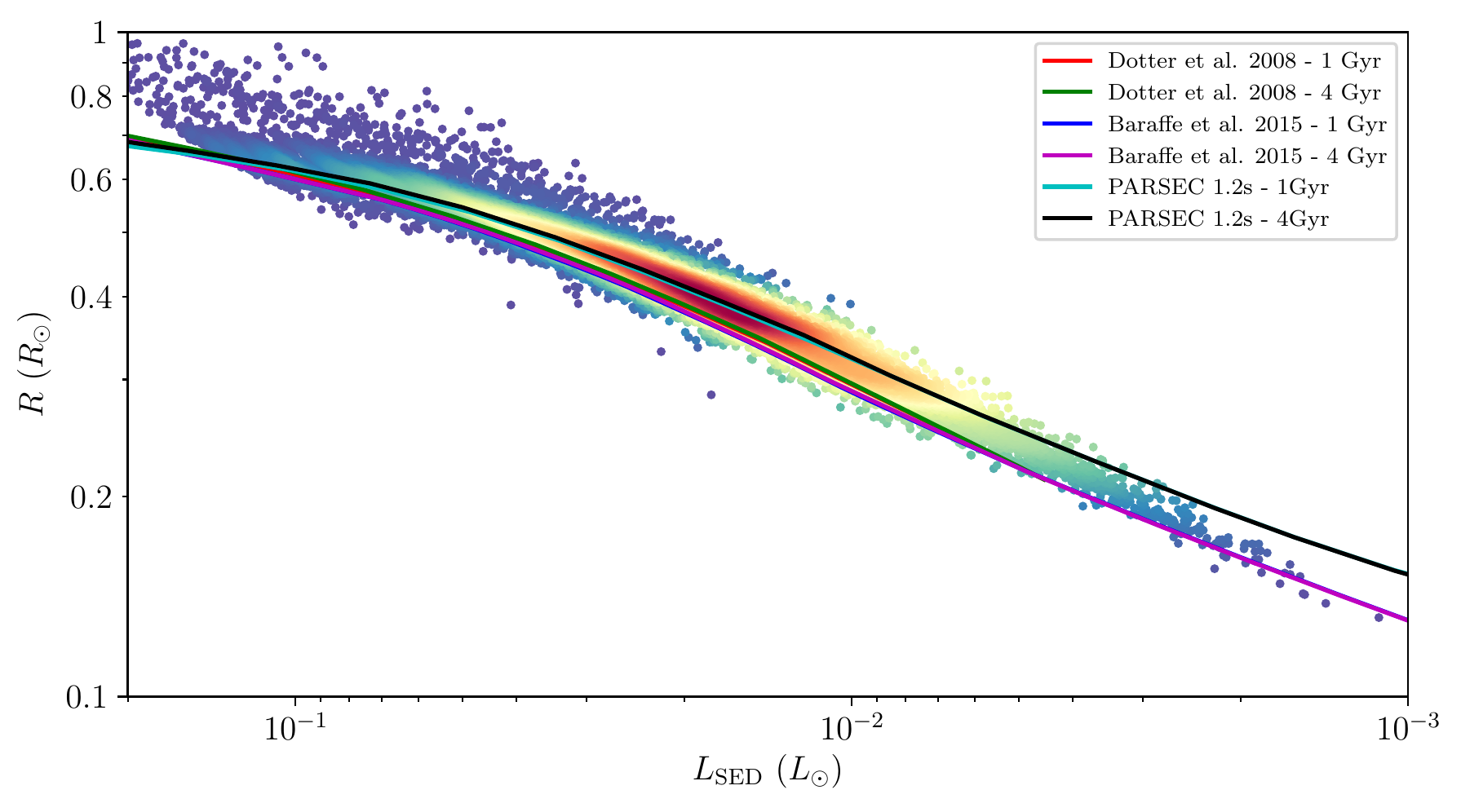}
    \caption{The \textbf{good} sample of our dataset in the $L_{\rm SED} - R$ plane. Due to the strong correlation between the axes in the $T_{\rm SED} - R$ plane, this plane is preferable for accurately measuring radius inflation in our sample. Accompanying the data are the same isochrones as in \autoref{fig:rad-teff-const-lum}. The \citet{Dotter:2008aa} and \citet{Baraffe:2015aa} isochrones trace a similar sequence in this space, resulting in the \citet{Dotter:2008aa} isochrones being hidden behind the \citet{Baraffe:2015aa} isochrones. }
    \label{fig:rad-lum-plot}
\end{figure*}

In \autoref{fig:rad-teff-const-lum} the purely theoretical models undershoot the median of the radius distribution for all M-dwarf stars within our sample.
However the PARSEC 1.2S model, which is calibrated by adopting an empirical $T - \tau$ relation derived from DEBs as the boundary condition for the stellar interiors, traces the median radius well for the whole sample. 
When inferring the radius inflation of a sample of stars, the choice of parameters is of vital importance. 
As \autoref{fig:uncer-contours} shows, the uncertainties in $T_{\rm SED}$ and $R$ are strongly correlated; one cannot simply trace upwards from the theoretical sequence to infer the inflation. 
As mass is most closely related to luminosity, the radius inflation should in fact be measured in the more fundamental $L_{\rm SED} - R$ plane, shown in \autoref{fig:rad-lum-plot}.
We measured the radius inflation for each of the models used in this work by picking a point of identical luminosity from the models and finding the difference between the prescribed model radius and our inferred radius. 

\subsection{Temperature - Radius Relation} 
\label{sub:teff-r-relation}
\begin{figure*}
	\includegraphics[width=\textwidth]{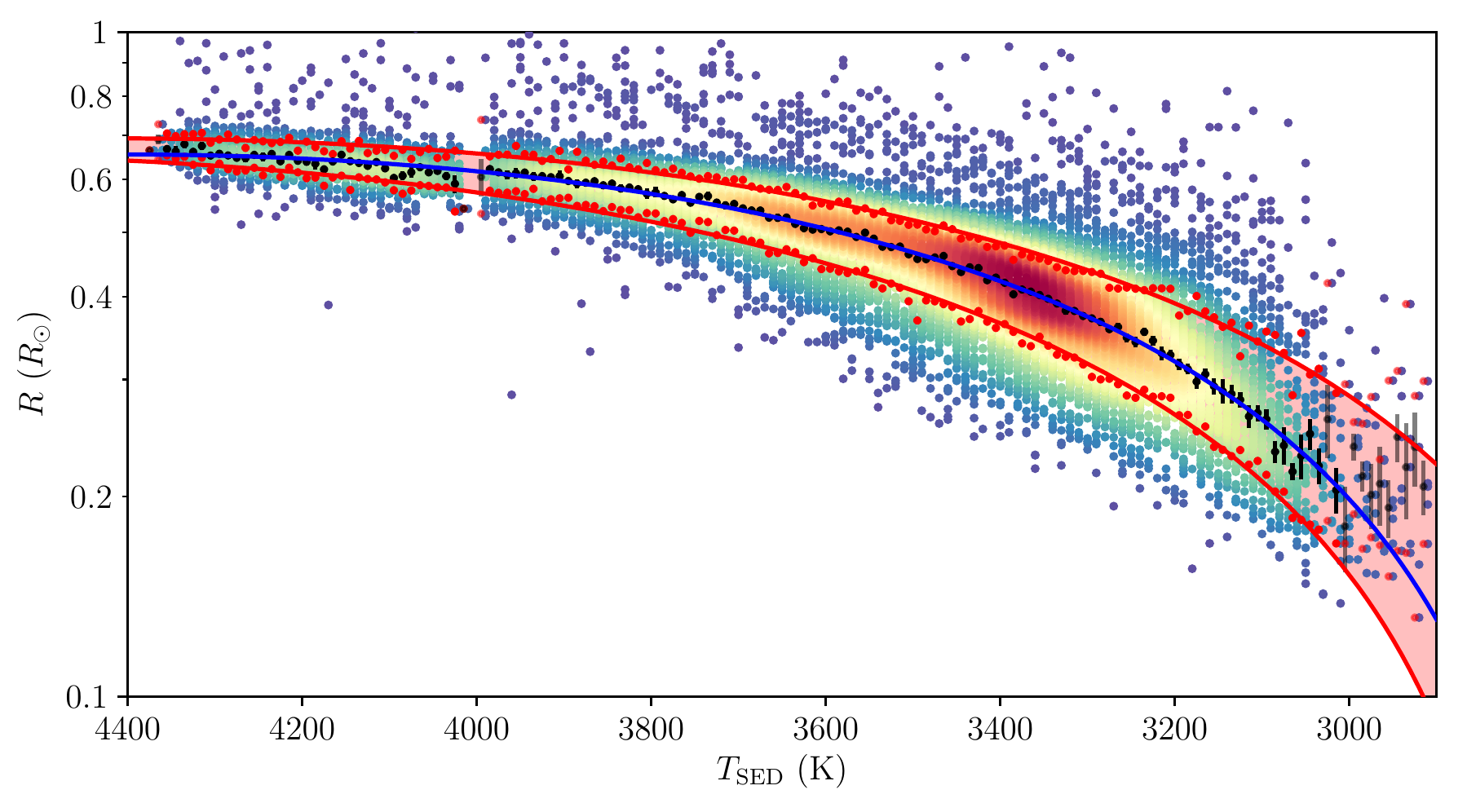}
	\caption{The $T_{\rm SED} - R$ relationship derived for our sample. The final relationship, given by \autoref{eq:r-teff-relation}, is the solid blue line along with its 68\% confidence intervals shown in red. The black dots show the stars used to perform the fit of the relationship. The red points bordering the upper and lower bound lines also show the stars used to fit them. The points in bins flagged as not good, and thus not used during fitting, are shown with semi-transparent markers. }
	\label{fig:relationship}
\end{figure*}

We can use our fits to derive a relation between temperature $T_{\rm SED}$ and radius $R$. 
We split the sample into 10K bins and took the median value from each bin. 
We required a minimum of 11 sources per bin, otherwise the entire bin was ignored. 
We fitted the medians with a $2^{\mathrm{nd}}$-order polynomial, which is shown in \autoref{fig:relationship} as the blue line, with the black points being the median values. 
The error bars are the standard deviation of the radius distribution within each bin. 
The relation that we fit from this sample is
\begin{align}
	R_{\rm fit}(T_{\rm SED})\ = &-3.861 + 2.054\times10^{-3}\ T_{\rm SED} -\ 2.335\times10^{-7}\ T_{\rm SED}^2, \nonumber\\ 
	&3000{\rm K} \le T_{\rm SED} \le 4400{\rm K}.
	\label{eq:r-teff-relation}
\end{align}
Using the same bins as before, we also drew the points that lie at the $16^\mathrm{th}$ and $84^\mathrm{th}$ percentiles to find the value of $1\sigma$ for each bin.
We used these points to fit further $2^{\mathrm{nd}}$-order polynomials to yield upper $R_\mathrm{high}$ and lower $R_\mathrm{low}$ confidence radii for each bin.
The upper and lower bound radii are given by
\begin{align}
	R_\mathrm{high}\ (T_{\rm SED}) = &-3.285 + 1.801\times10^{-3}\ T_{\rm SED} - 2.040 \times10^{-7}\ T_{\rm SED}^2,\nonumber\\& \quad 3000{\rm K} \le T_{\rm SED} \le 4400{\rm K},\label{eq:lower-bound-radii}\\
	R_\mathrm{low}\ (T_{\rm SED}) = &-3.288 + 1.691\times10^{-3}\ T_{\rm SED} - 1.813 \times10^{-7}\ T_{\rm SED}^2,\nonumber\\& \quad 3000{\rm K} \le T_{\rm SED} \le 4400{\rm K}.
	\label{eq:upper-bound-radii}
\end{align}
These functions are also shown in \autoref{fig:relationship}. 
These bounds are separated by $4\%$ at 4400K, increasing to $12\%$ at $3500$K and reaching a maximum separation of $30\%$ at the lower temperature limit of $3000$K. 
This scatter is further discussed in \autoref{sec:discussion}. 
The $T_{\rm SED} - R$ relation is tabulated in \autoref{tab:tsed-radius-relation}. 
\input{tab/tsed-radius-relation-table}

\subsection{Luminosity - Radius Relation}
\label{sub:rad-lum-relation}
\begin{figure*}
    \centering
    \includegraphics[width=\textwidth]{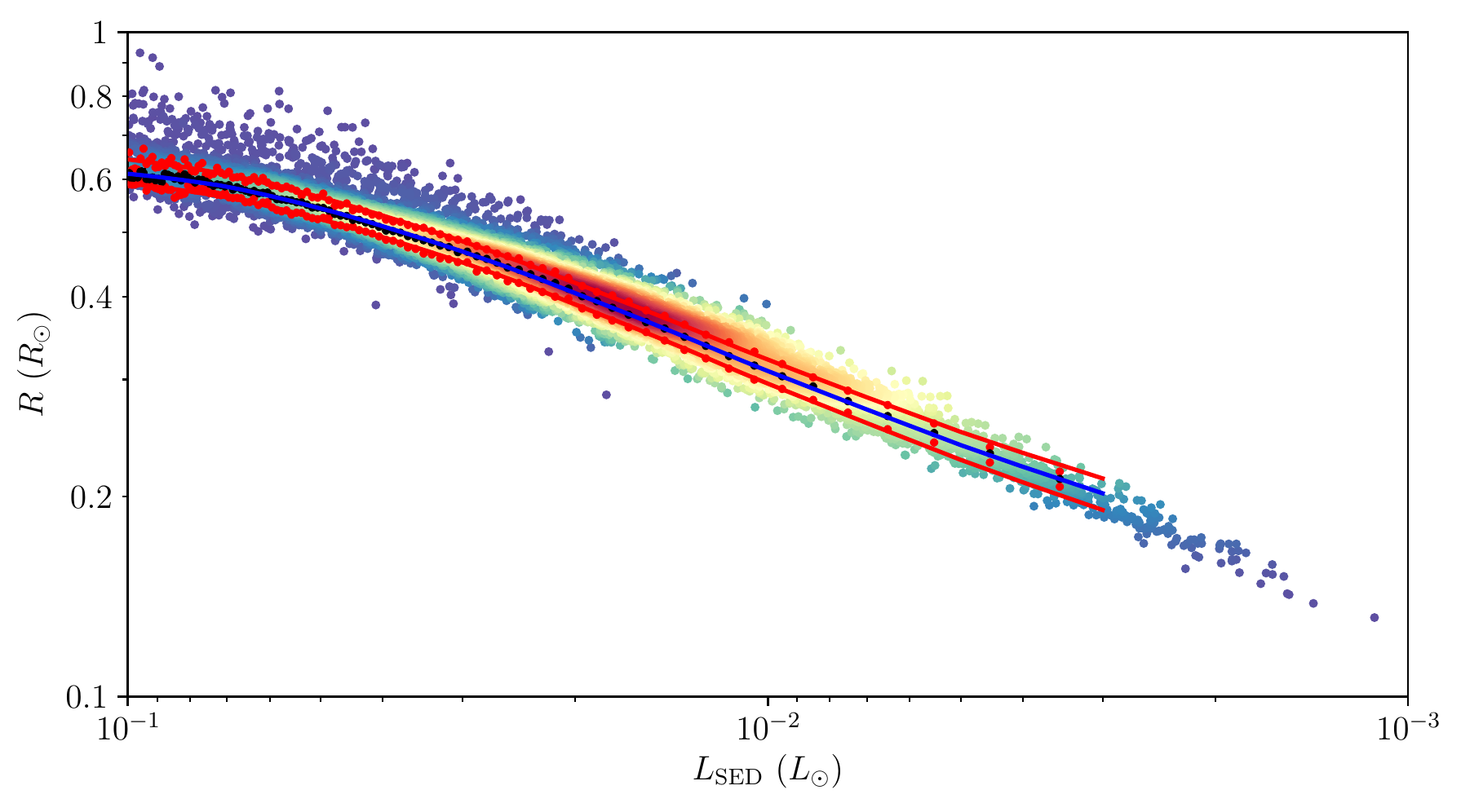}
    \caption{The luminosity - radius relationship plotted atop the stars from our sample. As with \autoref{fig:relationship}, we show the relation (blue), upper and lower bound lines (red) and the points from which each of the lines were fit in the corresponding colour. }
    \label{fig:rad-lum-corr}
\end{figure*}

\begin{figure}
	\includegraphics[width=\columnwidth]{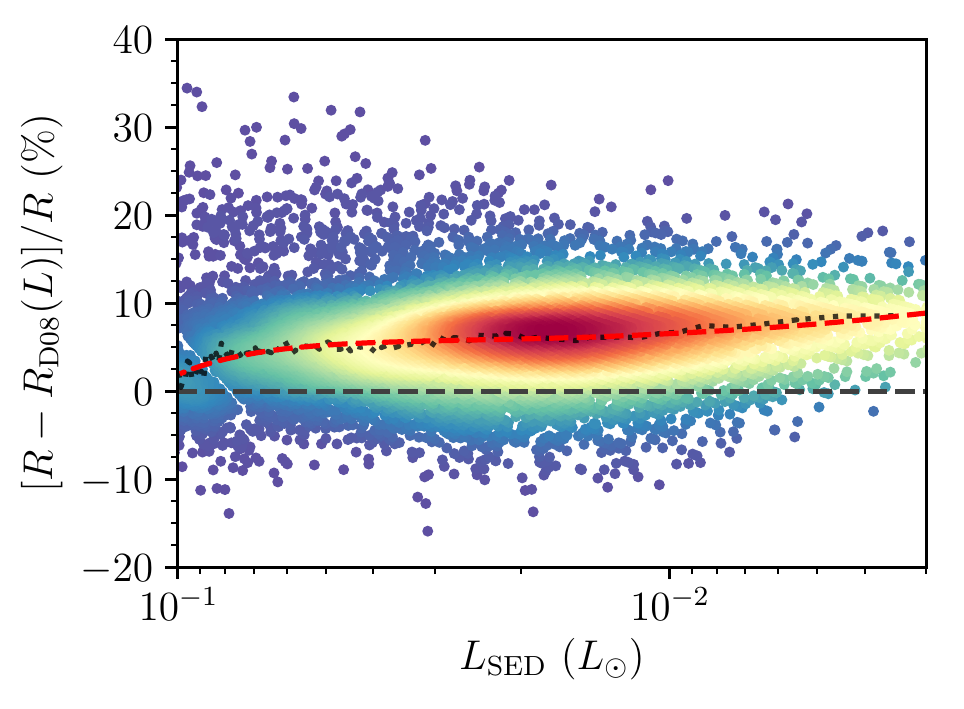}
	\caption{The radius inflation of our data from the 4 Gyr \citet{Dotter:2008aa} solar metallicity isochrone $[R - R_{\rm D08}(L)] / R$. The median radius inflation within each luminosity bin is shown as a black point. The luminosity - radius correction to this same isochrone is overlaid as a dashed red line. }
	\label{fig:rad-lum-spread}
\end{figure}

Temperature - radius relations are useful for the purposes of exoplanet host characterisation, however the stellar modelling community relies on more fundamental parameters when testing models. 
We have therefore transformed our temperature - radius data into the arguably more fundamental luminosity - radius plane, and used it to fit a relationship using a similar methodology. 
However, deriving a relation between luminosity $L$ and radius $R$ is more problematic, as high order polynomials are required to capture the detail in the relation. 
Despite falling below the majority of the radius distribution, the isochrones do a good job of predicting the shape of this dataset; suggesting the models capture the physical changes involved. 
Hence we created our $L_{\rm SED}$ - $R$ relationship as corrections to the \citet{Dotter:2008aa} 4Gyr solar metallicity isochrone. 
We subtracted the radius given by the isochrone from the median, upper and lower bound radius in each bin, leaving the difference between theoretical and observed radii. 
Then, to get the relation, we simply added our correction to the radius prescribed by the isochrone. 
This relation holds for values between $L_{\rm SED} = 0.003 L_{\odot}$ and $L_{\rm SED} = 0.1 L_{\odot}$. 
The correction to the \citet{Dotter:2008aa} isochrone is given by
\begin{align}
    R_{\rm fit}(L_{\rm SED})\ =\ &R_{\textrm D08}(L_{\rm SED}) + 0.0157 + 0.5670 L_{\rm SED} - 6.0633 L_{\rm SED}^2, \nonumber\\ 
	&0.003 L_{\odot} \le L_{\rm SED} \le 0.1 L_{\odot}
	\label{eq:dotter-rad-lum-correction}
\end{align}
where $L_{\rm SED}$ is the luminosity derived from our SED fitting and $R_{\textrm D08}(L_{\rm SED})$ is the theoretical radius of the star predicted by the \citet{Dotter:2008aa} isochrone at the given luminosity. 
The upper and lower bounds for the relation are given by 
\begin{align}
	R_\mathrm{high}\ (L_{\rm SED}) =\ &R_\textrm{D08}(L_{\rm SED}) + 0.0294 + 0.6833 L_{\rm SED} - 4.5443 L_{\rm SED}^2, \nonumber\\
	&0.003 L_{\odot} \le L_{\rm SED} \le 0.1 L_{\odot} \label{eq:lower-bound-radii-lum-dotter}\\
	R_\mathrm{low}\ (L_{\rm SED}) =\ &R_\textrm{D08}(L_{\rm SED}) + 0.0033 + 0.2634 L_{\rm SED} - 4.0079 L_{\rm SED}^2, \nonumber\\ 
	&0.003 L_{\odot} \le L_{\rm SED} \le 0.1 L_{\odot}.
	\label{eq:upper-bound-radii-lum-dotter}
\end{align}
The correction and final relation are shown atop the data in \autoref{fig:rad-lum-spread} and \autoref{fig:rad-lum-corr} respectively. 
The $L_{\rm SED} - R$ correction and the sum of the \citet{Dotter:2008aa} isochronal radius with the correction are tabulated in \autoref{tab:lsed-radius-relation}. 
\input{tab/lsed-radius-relation-table}

\section{Discussion}
\label{sec:discussion}

\subsection{Comparison with Literature Radii}
\label{sub:comparison-with-literature}
\begin{figure*}
    \centering
    \includegraphics[width=\textwidth]{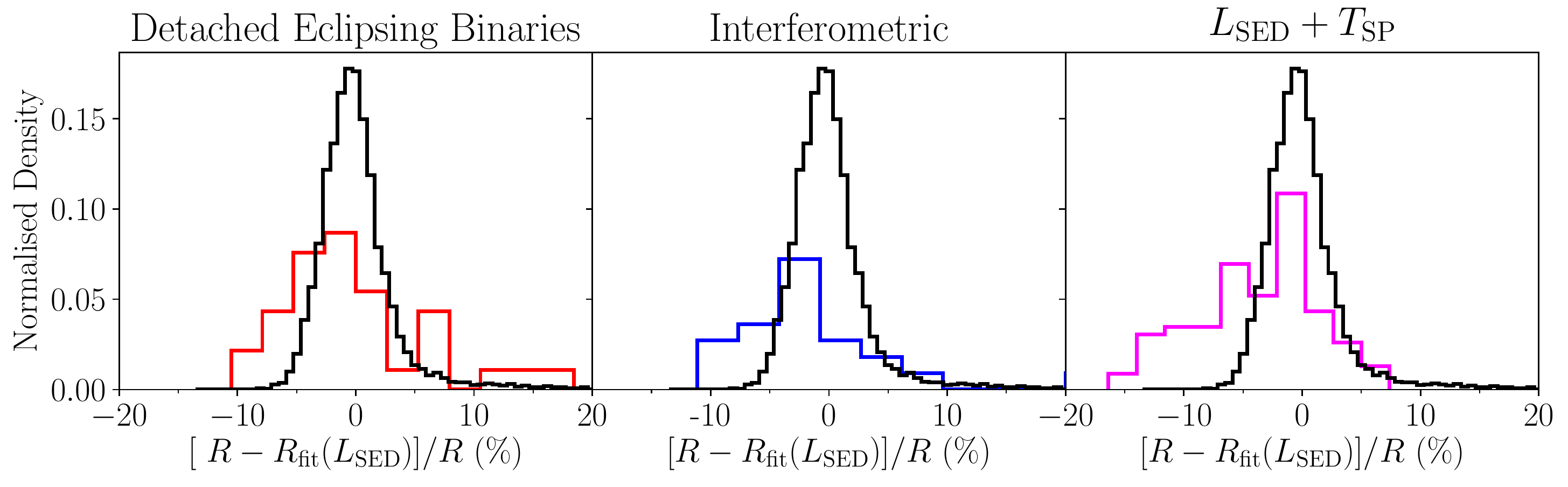}
    \caption{A comparison between the distributions of the relative residual of our measured radius $R$ with respect to our $L_{\rm SED} - R$ relation $R_{\rm fit}(L_{\rm SED})$; defined in \autoref{eq:dotter-rad-lum-correction}. Our sample is shown in black, with each of the others overplotted. Both the interferometric and DEB sample have median values about 3\% below our own. The \citet{Mann:2015aa} sample exhibits a bimodal structure, with the more pronounced peak occurring close to our median. }
    \label{fig:distribution-comparison}
\end{figure*}

To compare our measure of radius inflation with that from DEBs, interferometry and $L_{\rm SED} + T_{\rm sp}$ \citep{Mann:2015aa} we limited ourselves to the range $3400$K - $4400$K, where all methods are well sampled. 
The mode of all methods coincides at between $3 - 7\%$ inflated compared to the models.
In \autoref{fig:distribution-comparison} we plotted the distribution of relative radius residual from \autoref{eq:dotter-rad-lum-correction} (see \autoref{sub:rad-lum-relation}). 
The DEB and interferometric samples show medians at around $3\%$ lower than ours. 
The $L_{\rm SED} + T_{\rm sp}$ median corresponds well with our own, however it does exhibit a second peak at $7\%$ under inflation, though it is difficult to be sure whether this is a genuine feature of the population. 
There is also a long tail of outliers on the high inflation wing of our distribution; we suggest that these are a small number of binaries that have leaked into our sample. 

\begin{figure*}
    \centering
    \includegraphics[width=\textwidth]{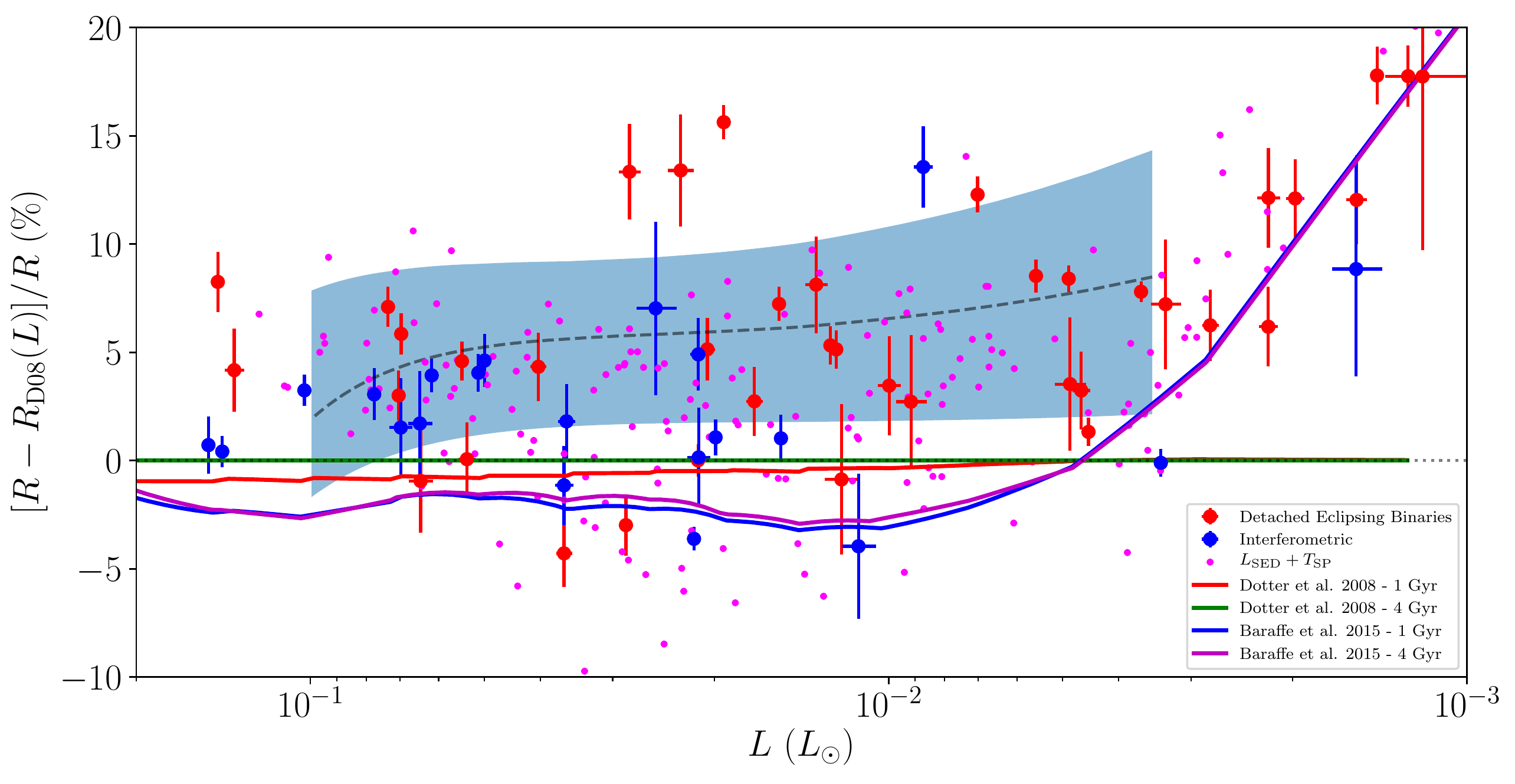}
    \caption{The radius inflation with respect to the \citet{Dotter:2008aa} 4 Gyr isochrone ($[R - R_{\rm D08}(L)] / R$) obtained by the four different methods. Inflation is plotted as a function of luminosity. The red points show detached eclipsing binaries \citep{Southworth:2015aa,Parsons:2018aa}. The blue points show interferometric stars \citep{Boyajian:2012aa}. The pink points show stars from \citet{Mann:2015aa}. The error bars for this sample have been omitted for reasons of clarity. The luminosity - radius relation derived in this paper (grey dashed line) with its associated $68\%$ density bounds is also shown. For comparison, model isochrones for 1Gyr and 4Gyr ages are shown for \citet{Dotter:2008aa} at $z=0.018$ and \citet{Baraffe:2015aa}. The original sources for \citep{Southworth:2015aa} are \citet{Welsh:2015aa,Torres:2014aa,Torres:2002aa,Terrien:2012aa,Stassun:2004aa,Rozyczka:2009aa,Rozyczka:2007aa,Pietrzynski:2013aa,Morales:2009ab,Morales:2009aa,Kraus:2017aa,Kraus:2011aa,Heminiak:2015aa,Hartman:2018aa}. The original sources for \citet{Parsons:2018aa} are \citet{Pyrzas:2012aa,Parsons:2018aa,Parsons:2016aa,Parsons:2012ac,Parsons:2012ab,Parsons:2012aa,Parsons:2010aa}. }
    \label{fig:conclusion-comp-plot}
\end{figure*}

In \autoref{fig:conclusion-comp-plot} we compare the interferometric, eclipsing binary and $L_{\rm SED}-T_{\rm sp}$ datasets with our own as a function of luminosity, as we require that the datasets have another physical quantity in common in addition to the radius.
For the reasons outlined in \autoref{sub:discussion-radius-inflation} luminosity is the best abscissa to use. 
Although the eclipsing binaries are normally viewed as a mass-radius dataset, the eclipsing binaries to which we are comparing also have temperatures derived from the spectra or photometric surface brightness, which in combination with the radius allows us to calculate a luminosity. 
The natural plane for the interferometric data is the luminosity-radius plane. 
The $L_{\rm SED} + T_{\rm sp}$ dataset can again be converted into $L_{\rm SED}$-radius.
The data presented in this paper are derived in the $T_{\rm SED}$-radius plane, and so can be converted into the $L_{\rm SED} - R$ plane.
We emphasize that all of these comparisons can only be made assuming $T_{\rm SED}\ =\ T_{\rm sp}\ =\ T_{\rm eff}\ =\ T_{\rm br}$, where $T_{\rm br}$ is the brightness temperature measured for some eclipsing binaries.
As \autoref{fig:conclusion-comp-plot} shows, all datasets show a radius inflation with respect to the models. 

For the luminosity range for which our $L_{\rm SED}-R$ relation is valid, there are a total of 29 DEBs, 16 interferometric stars and 154 $L_{\rm SED} + T_{\rm sp}$ stars. 
We determined which side of our median radius these samples fall. 
We found the $66\%$ of the DEBs, $88\%$ of the interferometric stars and $80\%$ of the $L_{\rm SED} + T_{\rm sp}$ sample lie below our median; meaning that all 3 other methods yield a lower median radius than ours. 
This skew to lower radii across this range supports the distributions shown in \autoref{fig:distribution-comparison}. 

This difference between the methods might be due to starspots.
First, any method which measures a spectroscopic temperature, especially in the optical, will not be sensitive to spots, because the immaculate photosphere spectrum will dominate the spot spectrum, leading to an over-prediction of the $T_{\rm eff}$ and under-prediction of radius. 
This affects both the eclipsing binary and $L_{\rm SED} + T_{\rm sp}$ sample and would result in radii smaller than those measured by our method. 
Although the $L_{\rm SED} + T_{\rm sp}$ and interferometric methods integrate under the SED of the star to find the luminosity, as we do, many of their objects saturate in WISE and so lack mid-IR coverage.
Not sampling the region of the SED where the spotted photosphere has its strongest contribution relative to the immaculate photosphere, and may make a non-negligible contribution to the overall stellar flux, may cause the measured luminosity to be too low, again resulting in radii that are too small. 
In summary, we found the radii measured by all methods to be inflated above the theoretical sequence, albeit we measure radii that are larger than the other methods.  

\subsection{Contributions to the Radius Scatter}
\label{sub:discussion-radius-inflation}
\autoref{fig:conclusion-comp-plot} demonstrates that for a given luminosity there is a $3 - 7\%$ spread in measured radius. 
It is important to determine whether this radius spread is real, and if so what effects contribute to it. 

\subsubsection{Could observational uncertainties contribute to the scatter?}
\label{sub:obs-uncertainties-scatter}
We first wished to establish whether the uncertainties in our radius determination could explain the spread before searching for a physical origin. 
\autoref{fig:uncer-contours} shows $68\%$ confidence contours for the uncertainty in $R$ and $T_{\rm SED}$ for an unbiased sub-sample of our catalogue, which we discussed in \autoref{sub:photometric_fitting}. 
The mean uncertainty in this selection of stars is $(1.6\%)$; much less than the $3 - 7\%$ spread that is observed. 
Furthermore we find a similar (but slightly larger) spread in the literature radii, with $62\%$ of the DEBs, $50\%$ of interferometric stars and $53\%$ of the $L_{\rm SED} + T_{\rm sp}$ sample lying within our relation's $68\%$ confidence bounds.
Thus we conclude that the spread is not the result of observational uncertainties.

\subsubsection{Does flux contamination from faint counterparts contribute to the scatter?}
\label{sub:flux-contamination}
\citet{Wilson:2017aa} show that the AllWISE bands can suffer contamination due to faint, hidden sources falling within the large PSFs of brighter stars. 
As well as making accurate catalogue matching problematic in crowded fields, this can also cause contamination to AllWISE photometry. 
Even a modest flux contamination from a stray faint source within the WISE PSF has the potential to cause a large discrepancy in both the retrieved $T_{\rm SED}$ and $R$. 
Fortunately, \citet{Wilson:2018aa} provide a catalogue of Gaia DR2 - WISE matches in the galactic plane, which allows us to assess the effect of contamination on our sample. 
According to their work, of the 2\,334 sources that match between the catalogues, fewer than $4\%$ were likely to be contaminated by $10\%$ or more.
The rest of the sky is less prone to crowding, so we suggest that $4\%$ of sources suffering contamination is the upper limit for our entire sample. 
Were our AllWISE photometry affected by contamination due to the presence of an unseen counterpart, we would expect the contaminated sources to exhibit more inflated radii than clean sources. 
However, we found no correlation between radius inflation and predicted flux contamination or probability; indicating that contamination in the WISE bands is unlikely to contribute towards the spread.

\subsubsection{Do starspots contribute towards the radius spread?}
\label{sub:starspots-spread}
Starspots introduce a second, cooler component to the SED; effectively diverting some of the luminosity of the star away from the immaculate photosphere. 
\citet{Jackson:2014aa} used a polytropic model including starspots to reproduce the radii of pre-main sequence (PMS) stars in the Pleiades and NGC 2516, which required spot coverages of 35 - 51\%. 
Using these models, they were able to find an 8\% inflation in the stars when compared to the \citet{Baraffe:2015aa} stellar models. 
\citet{Higl:2017aa} were also able to explain observed radii with starspots by covering large percentages (up to 44\%) of the surface of their models with spots. 

Our fitting assumed that the entire surface of the star is a single temperature. 
To examine the effect of starspots on our measured radii we produced a catalogue of simulated magnitudes of stars with spots and ran this catalogue through the same fitting process used for our observed sample. 
We simulated the input catalogue by sampling from a grid with $\log(g) = 5.0$ and varying spot filling factor $\gamma$ between $\gamma = 0.0 - 1.0$; whereas in all grids used for fitting data we have assumed $\gamma = 0.0$ (no spots). 
The composite photosphere consists of an immaculate and a spotted photosphere with temperatures $T_{\rm imac}$ and $T_{\rm spot}$ respectively. 
In determining the temperature of the spotted photosphere, we make the reasonable assumption \citep[see][]{Berdyugina:2005aa} that $T_{\rm spot} = 0.8\ T_{\rm imac}$ for $5000{\rm K} < T_{\rm imac} < 3000{\rm K}$. 
The synthetic magnitudes $Z_i$ in this grid are given by
\begin{equation}
    Z_{\lambda, {\rm syn}} = - 2.5 \log_{10} \left[ \frac{\int_\lambda \left( (1 - \gamma)  I_{\lambda {\rm imac}}  + \gamma I_{\lambda {\rm spot}}\right) S_{\lambda,i} d\lambda}{\int_\lambda f^\circ_\lambda S_{\lambda,i} d\lambda} \right] + m^\circ_\lambda,
    \label{eq:spotted-zlamsyn}
\end{equation}
where $I_{\lambda {\rm imac}}$ and $I_{\lambda {\rm spot}}$ are the intensity of the immaculate and spotted photosphere respectively.
The effective temperature $T_{\rm eff}$ of these spotted models becomes
\begin{equation}
    T_{\rm eff, spotted} = \left( (1 - \gamma)\ T_{\rm imac}^4 + \ \gamma\ T_{\rm spot}^4 \right)^{\frac{1}{4}}. 
    \label{eq:teff-spotted}
\end{equation}

To produce the simulated input catalogue we iterated through immaculate photosphere temperature, adopting the effective temperature of the combined photosphere as given in \autoref{eq:teff-spotted}. 
Each $T_{\rm eff, spotted}$ was mapped onto the corresponding stellar radius given by \autoref{eq:r-teff-relation}. 
We utilised a Monte Carlo method to account for the uncertainties, which could potentially add to the spread.
For each band we generated a CDF for the uncertainties in the observed catalogue and Monte Carlo sampled it for each simulated star. 
With this catalogue we then performed a fitting using the unspotted grid. 

\begin{figure*}
    \centering
    \includegraphics[width=\textwidth]{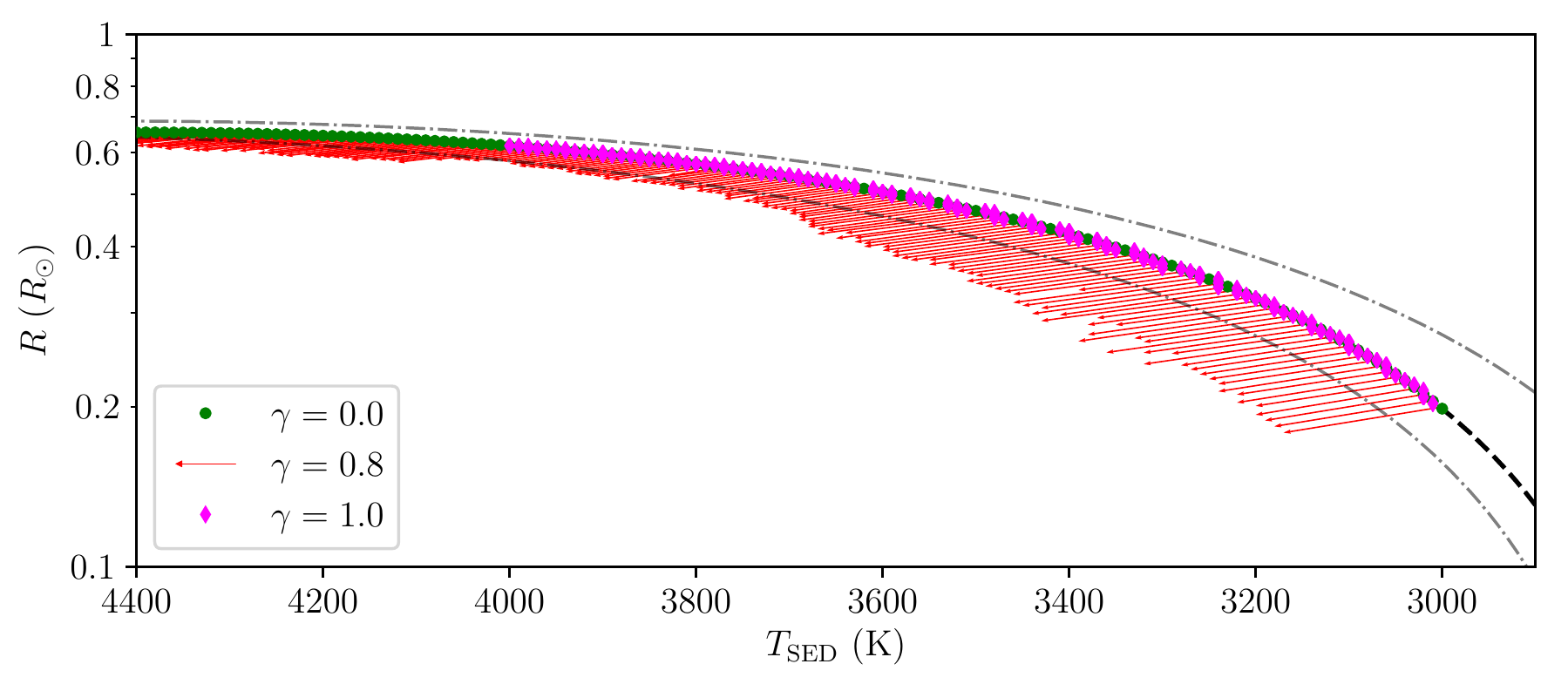}  
    \caption{The points show the radius and temperature retrieved by our fitting a catalogue of simulated synthetic photometry with varying spot filling factor $\gamma$. We show 0\% ($\gamma = 0.0$) and 100\% ($\gamma = 1.0$) spot coverage, which lie along the relation, along with $\gamma = 0.8$ which lies at the extremity of the spread. We show the 68\% density bounds of our $T_{\rm SED} - R$ relation as grey dotted lines. }
    \label{fig:spot-spread-plot}
\end{figure*}

The fits resulting from this process is shown in \autoref{fig:spot-spread-plot}. 
What this makes clear is that spot coverage can contribute towards the perceived spread in radius. 
At $\gamma = 0.0$ and $\gamma = 1.0$, we recover the original $T_{\rm SED} - R$ relation as would be expected. 
For $0.0 < \gamma \le 0.8$ the stars scatter to lower radii at increased $T_{\rm SED}$, conserving the overall luminosity. 
However, for $ 0.8 < \gamma < 1.0$ we see the effects of the spotted photosphere in the SED and the fits begin retrieving temperatures closer to those of the spots and the measured radii become closer to the relation.
This has the effect of producing a scatter in our relation which closely corresponds to the $68\%$ confidence lines resulting from our $T_{\rm SED} - R$ relation presented in \autoref{sub:teff-r-relation}.

\subsubsection{Correlations with activity}
\label{sec:activity}
\begin{figure*}
    \centering
    \includegraphics[width=\textwidth]{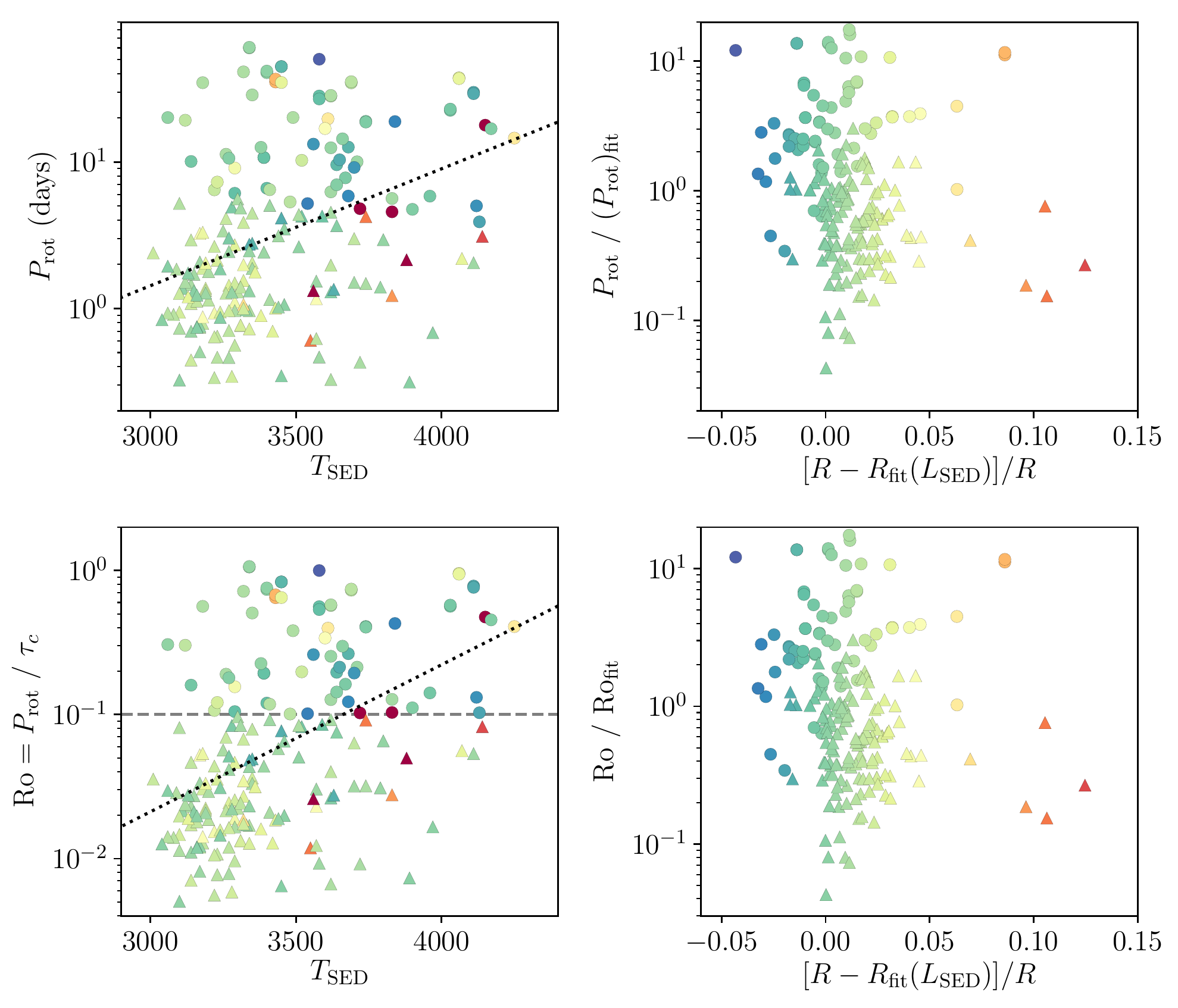}
    \caption{The correlation between our measured relative radius residual $[R - R(L)] / R$ and rotation period $P_{\rm rot}$ (left) and Rossby number ${\rm Ro}$ (right). The sample is divided between saturated (triangles) and unsaturated (circles) activity samples. The transition is marked by a grey dashed line and determined to be ${\rm Ro} = 0.1$, both observationally \citep{Newton:2017aa} and theoretically \citep{Reiners:2009aa}. We removed spurious correlations with spectral type from the right pane by performing a linear fit, shown as a black dotted line, in the accompanying left pane. Several stars at extreme inflations are not visible in the plot. The colour map denotes $[R - R_{\rm fit}(L_{\rm SED})] / R$. }
    \label{fig:rotation-rossby-correlation}
\end{figure*}

Although we have shown that the effect of starspots on our measurement technique is able to explain our scatter, this hypothesis would necessitate a correlation between measured radius and magnetic activity. 
To probe magnetic field strength we checked for correlations between radius inflation and markers of magnetic activity, the most reliable of which is rotation period $P_{\rm rot}$. 
We investigated a correlation with rotation period by assembling a sample of periods from  \citet{McQuillan:2013aa} and \citet{McQuillan:2014aa} observed using Kepler. 
Unfortunately, there are only 21 targets in common with our sample, so we chose to supplement these catalogues with rotation periods determined from Gaia DR2 lightcurves in \citet{Lanzafame:2018aa} which are based on sparser lightcurves, but add an appreciable number of stars; resulting in a final sample of 189 stars for which we have rotation periods. 
We used a theoretical expression for convective turnover time $\tau_c$, provided by \citet{Cranmer:2011aa}, to determine the equivalent Rossby number ${\rm Ro} = P_{\rm rot} / \tau_c$  for each star. 
To avoid spurious correlations with spectral type, we first performed a linear fit on both $P_{\rm rot}$ and ${\rm Ro}$ vs $T_{\rm SED
}$ and corrected for it in our final correlations, which are shown alongside the fits in \autoref{fig:rotation-rossby-correlation}.
Rotation alone is adequate to show that our sample lacks appreciable correlations between magnetism and radius inflation, and is the most fundamental because more rapid rotation rates presumably promote a stronger dynamo action within the stellar interior. 
However, \citet{Wright:2011aa} and \citet{Newton:2017aa} show that X-ray luminosity and $H_\alpha$ excess correlate well with rotation period in their unsaturated regimes. When unsaturated both can be used as additional markers of magnetic fields emergent at the stellar surface. 
It has been shown both observationally \citep{Wright:2011aa, Newton:2017aa} and theoretically \citep{Reiners:2009aa} that activity saturates at around ${\rm Ro} \simeq 0.1$, meaning that $31\%$ of our rotation sample would be in the unsaturated regime. 
Although rotation itself does not saturate, we have split our rotation periods into saturated and unsaturated sub-samples around this threshold to aid comparison with the following samples, which do. 

\begin{figure*}
    \centering
    \includegraphics[width=\textwidth]{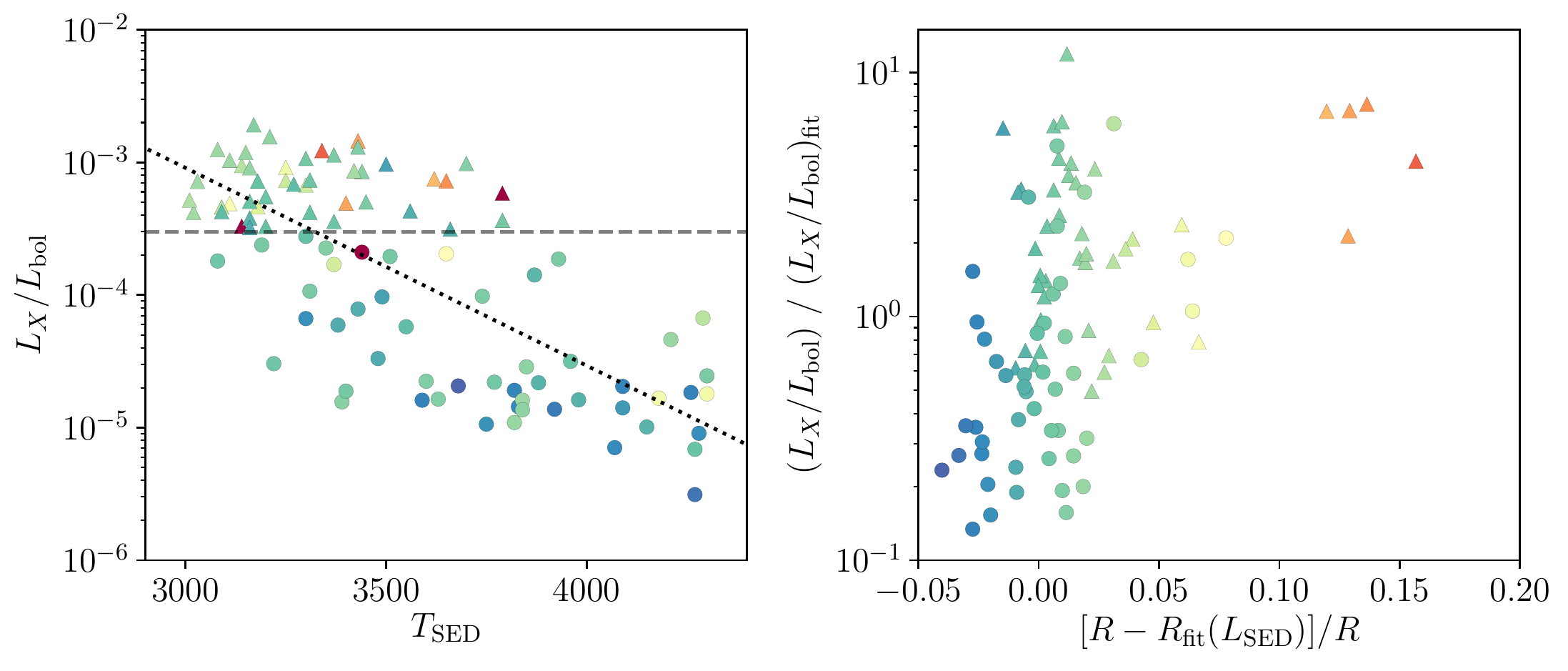}
    \caption{Our sample of stars with X-ray luminosity $L_X / L_{\rm bol}$ against relative radius residual $[R - R_{\rm fit}(L_{\rm SED})] / R$. The sample is divided into stars that lie in the saturated (triangles) and unsaturated (circles) regimes. This transitions is marked by a grey dashed line drawn at $L_X / L_{\rm bol} \simeq 3\times10^{-4}$, which corresponds to the lower limit from \citet{Wright:2018aa}. The left pane shows a spurious correlation in our sample between $L_X / L_{\rm bol}$ and $T_{\rm SED}$, which is corrected for in the right plot using a linear fit (black dotted line). Several stars are omitted from the high inflation wing of this plot. }
    \label{fig:radius-inf-xray}
\end{figure*}

We investigated the correlation with X-ray luminosity by crossmatching with DR6 of the XMM-Newton Serendipitous Source Catalog \citep[3XMM DR6; ][]{Rosen:2016aa}, yielding 95 stars.
This sample was divided into saturated and unsaturated sub-samples using the threshold $L_X / L_{\rm bol} \simeq 3\times10^{-4}$, defined from the lower limit of the spread around the saturated sample of \citet{Wright:2018aa}; meaning $53\%$ of our stars are unsaturated. 
To avoid a spurious correlation with spectral type, and ensure that we only probe excess X-ray emission due to activity, as before we performed a linear fit to the data in the $T_{\rm SED} - L_X / L_{\rm bol}$ plane and used it to correct the values of $L_X / L_{\rm bol}$.
Both the fit and final correlation with relative radius residual are shown in \autoref{fig:radius-inf-xray}. 

\begin{figure}
    \centering
    \includegraphics[width=\columnwidth]{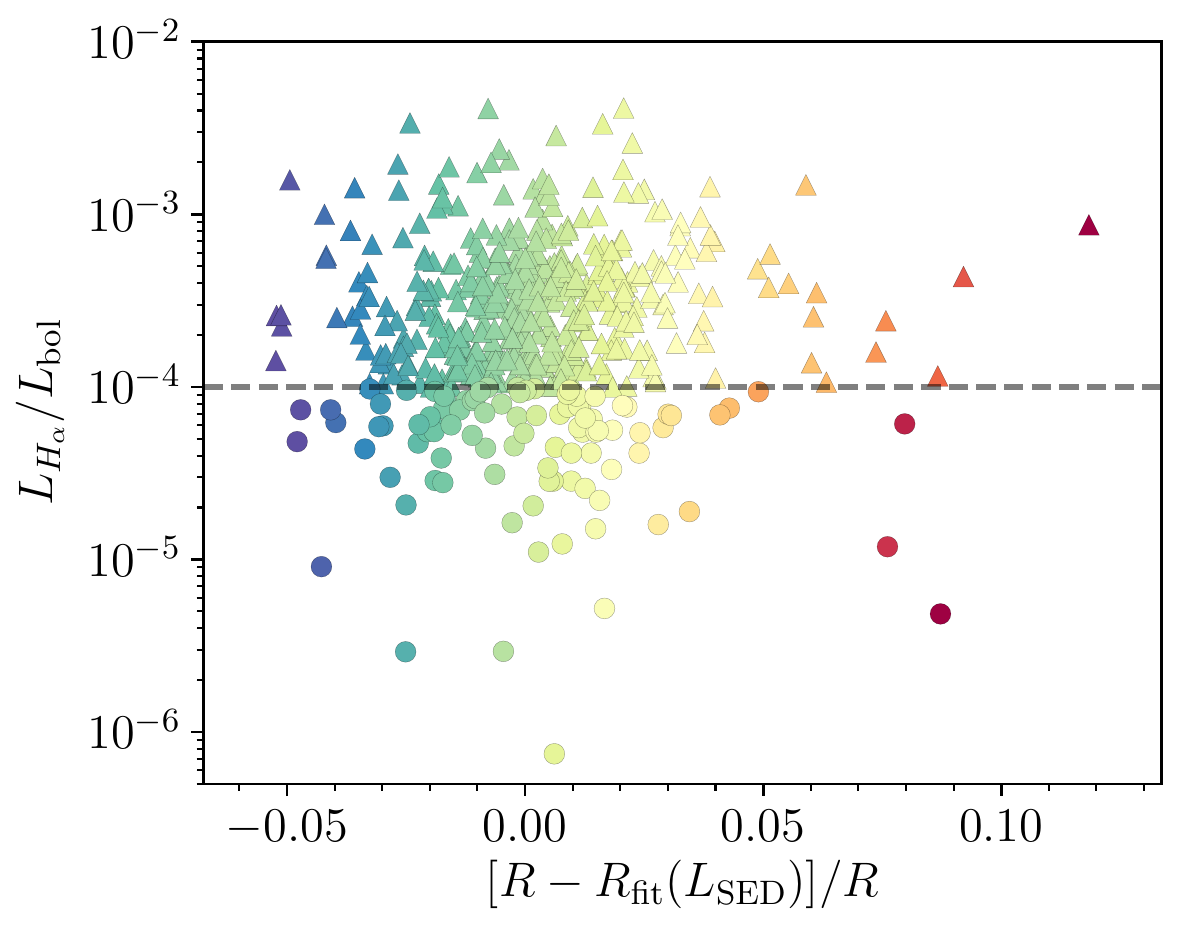}
    \caption{Our sample of stars for which we have $H_{\alpha}$ luminosity $L_{H_{\alpha}} / L_{\rm bol}$ against relative radius residual $[R - R_{\rm fit}(L_{\rm SED})] / R$. The sample is divided into stars that lie in the saturated (triangles) and unsaturated (circles) regime, delimited by the dashed line drawn at $L_{H_{\alpha}} / L_{\rm bol} = 10^{-4}$, corresponding to the value from \citet{Douglas:2014aa} and \citet{Newton:2017aa}. }
    \label{fig:radius-inf-halpha}
\end{figure}

Finally, we investigated the correlation with $H_\alpha$ by matching our catalogue with both DR2 of the INT Photometric H$\alpha$ Survey of the Northern Galactic Plane \citep[IPHAS2; ][]{Drew:2005aa,Barentsen:2014aa} and the VST Photometric H$\alpha$ Survey of the Southern Galactic Plane and Bulge \citep[VPHAS+; ][]{Drew:2014aa}. 
This results in a total of 573 stars for which we have $H_\alpha$ magnitudes. 
We have presented this sample in terms of $L_{H_\alpha} / L_{\rm bol}$ using
\begin{equation}
    \frac{L_{H_{\alpha}}}{L_{\rm bol}} = W_{H_\alpha} \chi,
    \label{eq:h-alpha-lum-ratio}
\end{equation}
where $\chi$, introduced in \citet{Walkowicz:2004aa}, was interpolated from the tabulated values from Table 8 of \citet{Douglas:2014aa}.
The equivalent width of the $H_\alpha$ line $W_{H_\alpha}$ due to activity was determined by measuring the excess flux across the $H_{\alpha}$ band, of width $\Delta \lambda_{H_{\alpha}}$, using 
\begin{equation}
    W_{H_\alpha} = \Delta \lambda_{H_{\alpha}} \left[ 10^{0.4\ (m_r - m_{H_\alpha})\ -\ (BC_{H_{\alpha}} - BC_r\textbf{})} - 1 \right],
    \label{eq:ha-equiv-width}
\end{equation}
where $m_r$ and $m_{H_{\alpha}}$ are the observed magnitudes, and $BC_r$ and $BC_{H_{\alpha}}$, the inactive model bolometric corrections.
This sample is shown in \autoref{fig:radius-inf-halpha}, along with the saturation threshold of $L_{H_{\alpha}} / L_{\rm bol} = 10^{-4}$ \citep{Newton:2017aa, Douglas:2014aa}. 
We note that some 53 of our $W_{H_{\alpha}}$ measurements are mildly negative, which indicates quiescence, thus we count these among our unsaturated sample. 
However due to resulting in negative $L_{H_\alpha} / L_{\rm bol}$ they do not appear on \autoref{fig:radius-inf-halpha}.
This demonstrates that our $H_\alpha$ sample spans both the saturated and unsaturated regimes, with around $25\%$ of our sample being unsaturated. 

To summarise, we have found that markers of both interior field strength ($P_{\rm rot}$) and surface field strength ($L_X / L_{\rm bol}$ and $L_{H_\alpha} / L_{\rm bol}$) show no appreciable correlation with radius inflation for M-dwarfs. 
All three markers are sampled in both the saturated and unsaturated activity regime, with between a quarter and a half of each being unsaturated. 

\subsubsection{How does metallicity affect the radius spread?} 
\label{sub:metallicity}
Differences in stellar metallicity could also cause a scatter in radii \citep{Berger:2006aa}. 
A reduction in metallicity, and thus opacity, would allow the star to more efficiently radiatively dissipate internal energy, resulting in a smaller radius at the same luminosity \citep{Berger:2006aa}. 
Stars within the solar neighbourhood show a metallicity spread with $\sigma = 0.2$ dex \citep{Boone:2006aa} with the lower extremity at $[M/H] \simeq - 0.6$ \citep{Neves:2013aa}. 
Comparing the \citet{Dotter:2008aa} solar metallicity isochrones with those of [M/H] = $\pm 0.25$ in luminosity - radius space, gives a difference in radius of about $\pm4\%$.  
So, theory suggests the contribution to the spread is minimal. 
However, \citet{Mann:2015aa} and \citet{Rabus:2019aa} show a correspondence between metallicity and relative residual in radius for a number of M-dwarf stars. 

\begin{figure}
	\includegraphics[width=\columnwidth]{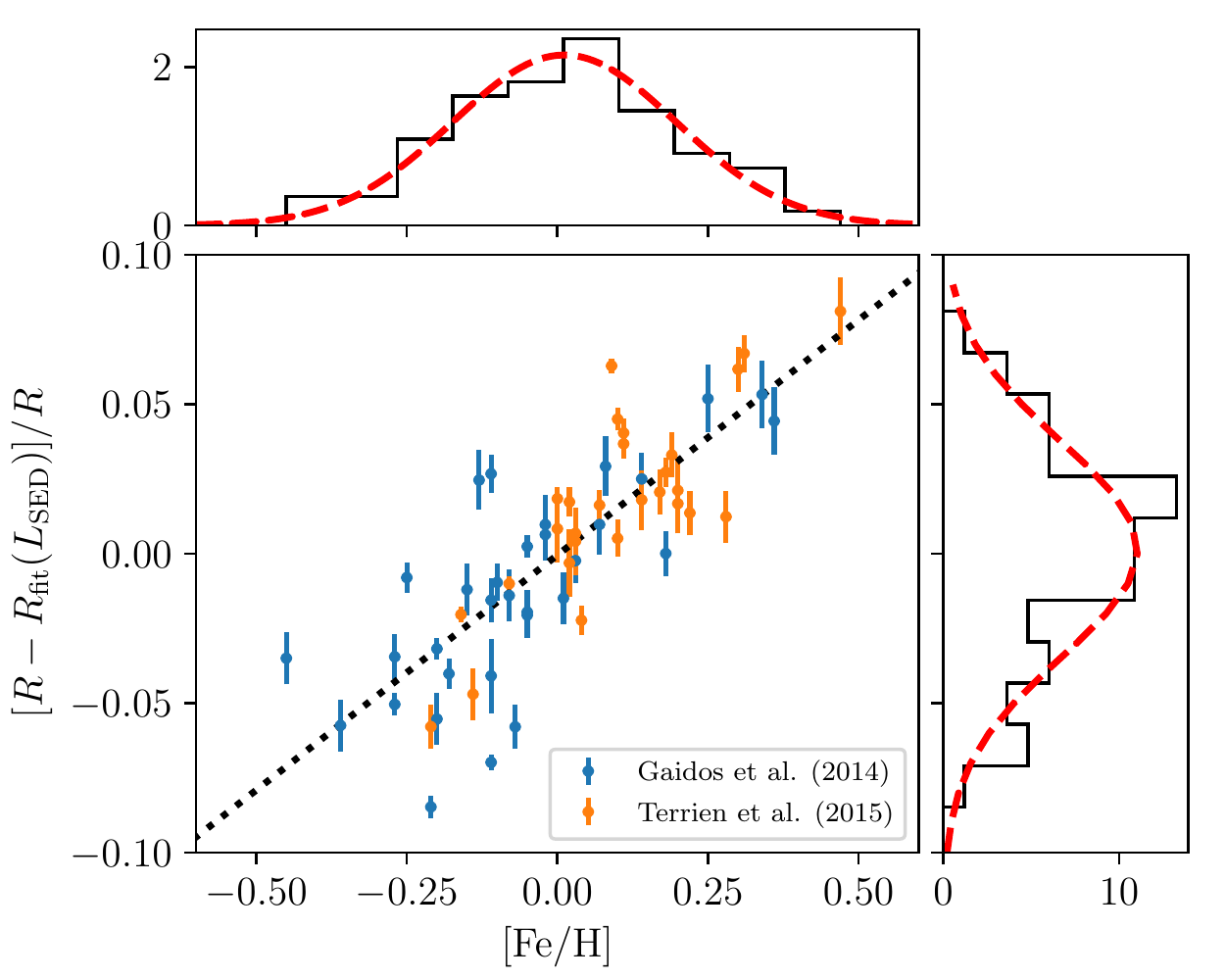}
	\caption{The correlation between radius residual and [Fe/H] for stars within our sample. The metallicities are from \citet{Terrien:2015aa} (orange) and \citet{Gaidos:2014aa} (blue). The radii and uncertainties are from our sample and fit using only solar metallicity atmospheres. }
	\label{fig:rad-feh-relation}
\end{figure}

To investigate this in our data, we assembled a sample of stars from our catalogue with measurements of [Fe/H] from \citet{Terrien:2015aa} and \citet{Gaidos:2014aa}. 
The sample with which we matched ranges between -0.5 $<$ [Fe/H] $<$ 0.5, indicating that our catalogue is relatively free of M-subdwarfs ([Fe/ H] $<$ -0.5). 
We verified this  using the Besan\c{c}on population synthesis model \citep{Robin:2003aa}, which yielded population sizes consistent with our sample and  indicated $\simeq 6\%$ of our sample has [M/H] $< -0.5$ and $\simeq 0.5\%$ having [M/H] $\le - 1.0$, as well as reproducing our distribution of metallicities well for [M/H > -0.5]. 
We ensured that these targets lie within the valid range of of the $L_{\rm SED} - R$ relation from \autoref{sub:rad-lum-relation} and we avoid the high inflation wing of our sample by choosing inflations $[R - R_{\rm D08}(L)]/R < 12\%$. 
We established in \autoref{sub:comparison-with-literature} that the radii from \citet{Mann:2015aa} are inconsistent with our own, so we chose not to use them.
The radius residuals resulting from the luminosity correction are shown as a function of [Fe/H] in \autoref{fig:rad-feh-relation}. 
The sample spans a residual of $\pm 6\%$, which appears to correlate strongly with metallicity, and corresponds well to the spread of metallicity in the solar neighbourhood. 

\begin{figure}
	\includegraphics[width=\columnwidth]{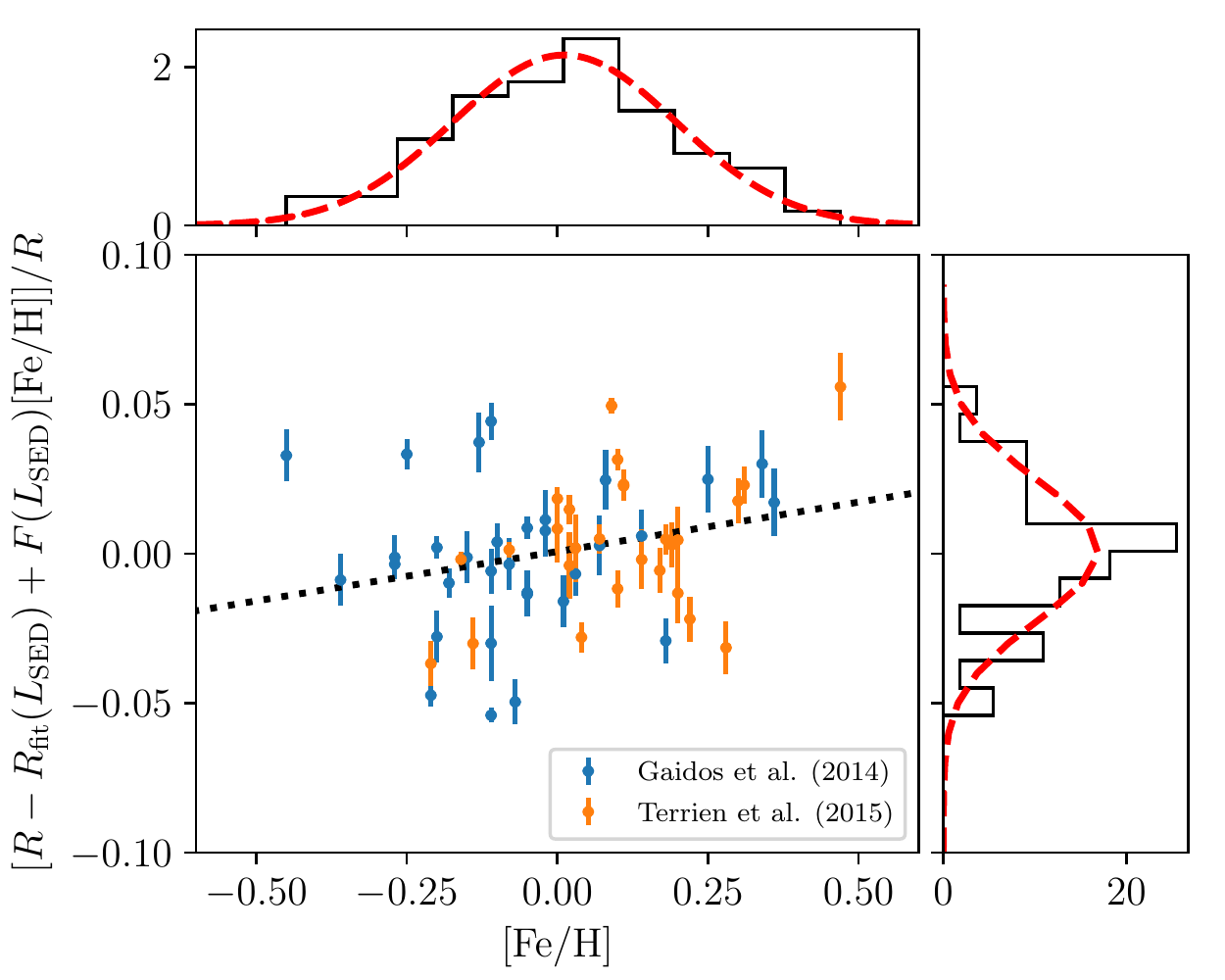}
	\caption{As \autoref{fig:rad-feh-relation} but with the correction from \autoref{sub:metallicity} applied. This correction accounts for using only solar metallicity atmospheres to fit a range of metallicities. }
	\label{fig:rad-feh-relation-flattened}
\end{figure}

This initially suggested that the spread was physical, and caused by metallicity, however the correlation is steeper than predicted by theory. 
Furthermore, we had measured all our radii using only solar metallicity atmospheres. 
To determine the error induced by this we produced grids for [M/H] $ = \pm 0.25 $ by interpolating synthetic photometry from [M/H] $ = \pm 0.5$ and [M/H] $ = 0.0$ grids\footnote{To ensure solar abundance for all metallicities we used the BT Settl AGSS2009 models, which adopts the \citet{Asplund:2009aa} solar abundances, as opposed CIFIST which uses those of \citet{Caffau:2011aa}. }\textsuperscript{, }\footnote{The BT Settl AGSS2009 model atmospheres are available from \href{https://phoenix.ens-lyon.fr/Grids/BT-Settl/AGSS2009/SPECTRA/}{https://phoenix.ens-lyon.fr/Grids/BT-Settl/AGSS2009/SPECTRA/}}.
We fitted each star in our input catalogue with all three metallicities, allowing us to produce a $L_{\rm SED} - R$ relation for each metallicity following the procedure from \autoref{sub:rad-lum-relation}. 
We found that our technique is able to determine luminosity consistently to within about $1\%$, regardless of which metallicity atmospheres are used. 
Thus, using the median radius from each luminosity bin, which we interpolated between bin midpoints, we calculated the difference between the radii measured at each metallicity and the radii measured at solar metallicity as a function of luminosity. 
Hence, we were able to measure the theoretical luminosity-dependent relationship between measured radius and [M/H] in the form of a linear relationship, with gradient $F(L_{\rm SED})$; values for which are tabulated in \autoref{tab:lum-feh-corr-grad}. 
This allowed us to determine the correct radius residual for each star, which are shown in \autoref{fig:rad-feh-relation-flattened}, using
\begin{equation}
    \frac{\delta R}{R} = \frac{1}{R}\left(R - R_{\rm fit}(L_{\rm SED}) + F(L_{\rm SED}) \left[ {\rm Fe}/{\rm H}\right] \right),
    \label{eq:rad-feh-corr}
\end{equation}
where $R_{\rm fit}(L_{\rm SED})$ is the $L_{\rm SED} - R$ relation given in \autoref{eq:dotter-rad-lum-correction}. 
\input{tab/lum-feh-corr}
The resulting corrected radii were fit with to yield another $L_{\rm SED} - R$ relation, which remains consistent to within $1\%$ with the relation presented in \autoref{sub:rad-lum-relation} for $0.015 L_{\odot} < L_{\rm SED} < 0.09 L_{\odot}$. 
When corrected for metallicity, we found that the correlation between relative radius residual and metallicity was no longer significant (see \autoref{fig:rad-feh-relation-flattened}). 

Therefore, the correlation of radius with metallicity in \autoref{fig:rad-feh-relation} is the result of fitting stars with a spread of metallicities with only solar metallicity atmospheres, which determines $T_{\rm SED}$ hence $R$ incorrectly. 
Thus the observed correlation between metallicity and radius spread in our sample is not physical and can be corrected for with accurate measurements of the stellar luminosity and metallicity. 
\citet{Mann:2015aa} also found a correlation between metallicity and radius within their sample, however it only appears as a function of $T_{\rm sp}$, not as a function of $M_{K_s}$ - a proxy for luminosity. 
Interestingly, the interferometric and DEB sample show a very similar scatter to our uncorrected radii, however we attribute this to observational uncertainty. 
The mean uncertainty in metallicity from \citet{Gaidos:2014aa} and \citet{Terrien:2012aa} will result in  a 1.7\% spread in our measured radius in \autoref{fig:rad-feh-relation-flattened}. 
Given that our median radius uncertainty from the fitting process is $1.6\%$ (see \autoref{sec:results}), this would be consistent with the $2.4\%$ scatter seen in \autoref{fig:rad-feh-relation-flattened}. 
There could also be an intrinsic scatter of the stars in our sample, but to remain consistent with the spread it would have to be at most $1 - 2 \%$. 
This leads us to conclude that our knowledge of M-dwarf radii is currently limited by the accuracy and precision of the metallicity measurement of the star. 

\subsubsection{The cause of the scatter - a summary}
\label{sub:the_cause_of_the_scatter}
We have been able to show that faint contaminants in AllWISE photometry were not the cause of our scatter (see \autoref{sub:flux-contamination}). 
We have also established that the uncertainties in our measurements were not able to explain the spread (see \autoref{sub:obs-uncertainties-scatter}).
We found that extremely high starspot coverage in most of the population could explain the scatter.
But since stars with small spot coverages would have lower measured radii, and stars with larger coverages, larger radii, this would imply a measured radius - activity correlation, which we were unable to find (see \autoref{sub:starspots-spread} and \autoref{sec:activity}). 
However, we found that there is a strong correlation between [Fe/H] and our measured radius. 
Although this could be misconstrued as a physical spread in radius caused by metallicity, we found that in fact this correlation resulted from using solar metallicity atmospheres to fit the SED of non-solar metallicity stars. 
When this is corrected for we found that the correlation disappeared, resulting in a scatter of about $2.4\%$ in the subset for which we could find metallicities. 
This spread is consistent with the $2 - 3\%$ found by \citet{Schweitzer:2019aa}. 
Uncertainty in [M/H] measurements accounts for about 1.7\% of this spread, with our uncertainties in radius explaining the remainder.  
The accuracy of the measured metallicities also makes determinations of effective temperature problematic, imposing another limit on the accuracy of our radii. 
Therefore we suggest that the spread of this corrected radius distribution is currently dominated by the accuracy and precision of current metallicity measurements (see \autoref{sub:metallicity}) and the intrinsic spread in M-dwarf radii is less than $1 - 2 \%$. 
This well characterised radius residual distribution also places an upper limit on the typical variability of spot filling factors of stars within our sample of $< 10\%$. 


\subsection{Explaining the Radius Inflation}
\label{sec:inflation-physics}
We now address the question as to why the M-dwarf main sequence is inflated from theoretical predictions, and note in passing the problem also appears to apply to the pre-main sequence \citep{Jackson:2018aa}. 
\autoref{sub:intro-causes-of-radius-inflation} summarises our current understanding of inflation - it is clear that the mechanism behind radius inflation in M-dwarfs remains a contentious point. 
So far the most compelling hypothesis has been dynamo driven magnetic fields, which would inhibit convection and probably modify the specific entropy in the convective region, and hence the internal structure of the star. 
Stars below a mass of approximately $0.35 M_{\odot}$ are thought to have interiors that are fully convective \citep{Limber:1958aa}, making this a satisfying explanation.
If this were to be the case, we might expect to see a spread in radius, from the theoretical non-magnetic models to some level of maximum inflation, provided we sampled a large range of rotation rates. 
However, we established in \autoref{sub:metallicity} that there is a tight (with perhaps a $1 - 2\%$ intrinsic scatter) main sequence for M-dwarf stars, refuting this scenario. 
One could argue that we have not adequately considered the case that radius inflation has saturated, much like the activity does, for all of the stars in our sample. 
\autoref{fig:rotation-rossby-correlation} shows we sample a large range in rotation period, including at least a quarter where activity is unsaturated. 
For saturation of radius inflation to be consistent with such tight a sequence, it  would have to occur at rotation rates much slower than both saturation of photospheric activity indicators (at ${\rm Ro} = 0.1$) and the slowest rotators in our sample ${\rm Ro \simeq 1.0}$. 

A further argument against stellar magnetism being responsible for radius inflation comes from the fact that that all of the indicators of magnetic activity we studied show no appreciable correlation with measured radius residual (see \autoref{sec:activity}). 
We would expect to see correlations between the activity and radius inflation were magnetism responsible. 
Given these two arguments, we reluctantly conclude that stellar magnetism is currently unable to explain radius inflation in main sequence M-dwarf stars. 

\subsubsection{Alternate explanations for radius inflation}
\label{sec:alternate-radius-inflation}
It is unlikely that magnetic fields are the cause of radius inflation in main sequence M-dwarf stars, so an alternate explanation must be sought. 
Metallicity is the main driving force behind the measured radius scatter, however in \autoref{sub:metallicity} it was shown to be caused by our measurement technique. 
This renders us unable to explain the inflated radii of main sequence M-dwarfs. 
However, one avenue of inquiry comes from the PARSEC 1.2S models \citep{Chen:2014aa,Marigo:2017aa} which adopts an empirical $T-\tau$ relation as the boundary condition to their interiors. 
As \autoref{fig:rad-teff-const-lum} and \autoref{fig:rad-lum-plot} show, this modification does a remarkably good job of characterising the degree of inflation and the intrinsic sequence for early to mid M-dwarf stars. 

\subsection{Determining Accurate M-dwarf Radii} 
\label{sec:determining_accurate_m_dwarf_radii}
We showed in \autoref{sub:metallicity} that metallicity can cause problems when determining M-dwarf radii. 
Using atmospheres of a different metallicity from the star being fitted causes the retrieved temperature, and thus the radius, to be incorrect. 
However, we have devised strategies for obtaining accurate M-dwarf radii in the face of these issues. 

\subsubsection{Without a metallicity measurement} 
\label{sub:radii_with_metallicity_measurement}
If there is not a measured metallicity for a star, there are two strategies available. 
Firstly, disregarding the metallicity of the star and fitting indiscriminately with solar metallicity models yields radii accurate to better than $5\%$; in our case about $3.6\%$. 
For statistical samples and less precise applications, this may be adequate. 
However, our method can measure the luminosity of the star correctly to within a couple percent regardless of metallicity (see \autoref{sub:metallicity}). 
Thus, we can use the measured luminosity along with an empirical $L_{\rm SED} - R$ relation, such as that in \autoref{sub:rad-lum-relation}, to obtain correct radii. The scatter about this relation ranges between $3.6\% - 4.5\%$. 

\subsubsection{With a metallicity measurement} 
\label{sub:radii_with_metallicity_measurements}
If the star has an accurate measurement of the metallicity, there are two more avenues open. 
The simplest option is to perform the fitting with an atmosphere of the appropriate metallicity.
However for large samples it can be impractical to generate large grids of synthetic photometry at a number of differing metallicities. 
So our other suggestion is to follow the method presented in \autoref{sub:metallicity}. 
This entails fitting the entire sample with solar metallicity models and correcting for the metallicity. 
Both strategies are highly dependent upon the accuracy and precision of the metallicity measurement. 
A metallicity constrained to about $10\%$ induces a scatter of about $1.7\%$ in measured radius. 
However, for cooler stars, with well constrained photometry and sufficiently accurate metallicities (better than about $3\%$), radius measurements of better than $1\%$ can be achieved with this method. 


\section{Conclusions} 
\label{sec:conclusions}
We have measured the temperature, radius and luminosity of a sample of 15\,274 late K and early M-dwarf stars using a modified spectral energy distribution fitting method. 
This method requires only accurate photometry and precision astrometry and thus adds a fourth method to those used to evaluate the veracity of stellar models.
Importantly, this method works natively in the $T_{\rm eff} - R$ space which is crucial for the characterisation of exoplanets.
We have derived empirical $T_{\rm SED} - R$ and $L_{\rm SED} - R$ relations, which can be used to characterise exoplanet host stars and validate stellar evolution models (\autoref{sub:teff-r-relation} and \autoref{sub:rad-lum-relation}). 

\noindent
The key conclusions of this work are as follows. 
\begin{enumerate}[i]
    \setlength{\itemindent}{1.5em} 

	\item Currently, none of the purely theoretical stellar models can describe the mean radius inflation of the main sequence at temperatures lower than about $4000$K. The measured radii are inflated by $3 - 7\%$ compared to those predicted by models (\autoref{sec:radius-inflation}). 

	\item We have shown that M-dwarfs lie on a tight sequence (with a scatter smaller than $1 - 2\%$) (see \autoref{sub:metallicity}). This is in conflict with magnetic models, which would suggest a spread in radius, from the theoretical sequence for non-magnetic models to a maximum inflation. 
	We have also shown that there is no appreciable correlation between all observational markers of magnetic activity and radius inflation (see \autoref{sec:activity}). This leads us to conclude that stellar magnetism is currently unable to explain radius inflation in main sequence M-dwarf stars (see \autoref{sub:discussion-radius-inflation}). Furthermore, this would explain the unexpected result that detached eclipsing binaries are not inflated with respect to their single star counterparts (see \autoref{fig:distribution-comparison}). 

	\item We discovered that fitting a distribution of metallicities with only solar metallicity models introduces an apparent correlation between [Fe/H] and $R$ (\autoref{sub:metallicity}). 

	\item However, we found that our technique correctly measures the luminosity regardless of metallicity, meaning that we can correct the measured radii. Without a measured metallicity we achieved a precision of $3.6\%$. However, this was improved to $2.4\%$ when corrected for metallicity. Given that the uncertainty in [M/H] accounts for $1.7\%$ of this spread, it is clear that the precision of metallicity measurements is currently the limiting factor for this method (\autoref{sec:determining_accurate_m_dwarf_radii}).
    
    \item In the absence of metallicity measurements we present an empirical $L_{\rm SED} - R$ relation which can be used to measure correct radii to a precision of $3.6 - 4.5 \%$ (see \autoref{sec:determining_accurate_m_dwarf_radii}). 
\end{enumerate}

What remains deeply puzzling is how tight the intrinsic sequence appears. 
Even if the stellar radii were identical at a given luminosity, we show in \autoref{sub:starspots-spread} that small changes in starspot filling factors would give an apparent spread in radius because of our measurement technique.
Hence the tight measured sequence implies that the starspot coverage among main sequence M-dwarfs is remarkably consistent, which is physically counter intuitive when considering the diverse levels of activity in even a modest subset of our full sample (see \autoref{sec:activity}). 

\section*{Acknowledgements}
SM is supported by a UK Science and Technology Facilities Council (STFC) studentship. 
The authors would like to thank Matthew Browning and Lewis Ireland for a critical reading of the manuscript. 
We would also like to thank Victor See and Sean Matt for useful conversations regarding stellar rotation. 
We are grateful to the referee for a very useful report, which sharpened our thinking and improved our presentation, in addition to providing useful background such as the discrepancies in binary parameters referred to in Section \ref{sec:measuring}.
Finally we would like to thank Stuart Littlefair for prompting us to investigate correlations with activity. 

This work has made use of data from the European Space Agency (ESA) mission {\it Gaia} (\url{https://www.cosmos.esa.int/gaia}), processed by the {\it Gaia} Data Processing and Analysis Consortium (DPAC, \url{https://www.cosmos.esa.int/web/gaia/dpac/consortium}). Funding for the DPAC has been provided by national institutions, in particular the institutions participating in the {\it Gaia} Multilateral Agreement.

This publication makes use of data products from the Two Micron All Sky Survey, which is a joint project of the University of Massachusetts and the Infrared Processing and Analysis Center / California Institute of Technology, funded by the National Aeronautics and Space Administration and the National Science Foundation.

This publication makes use of data products from the Wide-field Infrared Survey Explorer, which is a joint project of the University of California, Los Angeles, and the Jet Propulsion Laboratory/California Institute of Technology, funded by the National Aeronautics and Space Administration.

This paper makes use of data obtained as part of the INT Photometric H$\alpha$ Survey of the Northern Galactic Plane (IPHAS, www.iphas.org) carried out at the Isaac Newton Telescope (INT). The INT is operated on the island of La Palma by the Isaac Newton Group in the Spanish Observatorio del Roque de los Muchachos of the Instituto de Astrofisica de Canarias. All IPHAS data are processed by the Cambridge Astronomical Survey Unit, at the Institute of Astronomy in Cambridge. The bandmerged DR2 catalogue was assembled at the Centre for Astrophysics Research, University of Hertfordshire, supported by STFC grant ST/J001333/1.

Based on data products from observations made with ESO Telescopes at the La Silla Paranal Observatory under programme ID 177.D-3023, as part of the VST Photometric H$\alpha$ Survey of the Southern Galactic Plane and Bulge (VPHAS+, www.vphas.eu). 

This research has made use of data obtained from the 3XMM XMM-Newton serendipitous source catalogue compiled by the 10 institutes of the XMM-Newton Survey Science Centre selected by ESA.

This research has made use of the VizieR catalogue access tool, CDS, Strasbourg, France.
We gratefully acknowledge the software that made this work possible, including:
\textsc{STILTS} and \textsc{TOPCAT} \citep{Taylor:2006aa}, 
\textsc{Scipy} \citep{Jones:2001aa},
\textsc{Numpy} \citep{Oliphant:2015aa}, 
\textsc{Matplotlib} \citep{Hunter:2007aa} and
\textsc{Astropy} \citep{Astropy}. 

\bibliographystyle{mnras} 
\bibliography{ms-mdwarf-radii,deb-refs,software}

\label{lastpage}
\end{document}

%% file: tab/data-sources.tex
\begin{table*}
	\begin{tabular}{| c | c | c | c | c | c | c |}
	\hline
	Band & $\lambda_\text{iso}$ (\AA) & System & Selection Criteria & No. Sources & Coverage & Source \\
	\hline

	\multirow{2}{*} { -- } & \multirow{2}{*} { -- } & \multirow{2}{*} { -- } & \multirow{2}{*} { 
	\begin{tabular}{l}$r_\mathrm{est} \le 100 pc$\\ $(r_\mathrm{hi} - r_\mathrm{lo}) / r_\mathrm{est} < 0.01$\end{tabular}
	} & \multirow{2}{*} { 138\,279 } & \multirow{2}{*} { 100.0\% } & \multirow{2}{*} { \citet{Bailer-Jones:2018aa} } \\
	& & & & & & \\

	\hline
	\multirow{3}{*} { \begin{tabular}{l} $G_\mathrm{BP}$ \\ $G_\mathrm{RP}$ \end{tabular} } & 
	\multirow{3}{*} { \begin{tabular}{c} $5320$ \\ $7970$ \end{tabular} }  & 
	\multirow{3}{*} { Gaia DR2 } & \multirow{3}{*} {
	\begin{tabular}{c}
		$G_\mathrm{BP}$ - $G_\mathrm{RP}$\ >\ 1.5 \\
		$\pi > 9.0$ \\
		$\delta \pi / \pi < 0.02$
	\end{tabular}
	} &\multirow{3}{*} { 99\,288} & \multirow{3}{*} { 71.8\% }& \multirow{3}{*}{ \begin{tabular}{l} \citet{Gaia2016} \\ \citet{Gaia2018} \\ Passbands: \citet{Evans:2018aa} $^{\alpha, \beta}$ \end{tabular} \hfil }  \\
	& & & & & & \\
	& & & & & & \\
	





	\hline
	$J_\text{2MASS}$ & 12410 & \multirow{3}{*} { 2MASS } & \multirow{3}{*} { 
	\begin{tabular}{l}$\mathrm{ph\_qual} = \mathrm{"AAA"}$\\\end{tabular}
	} &\multirow{3}{*} { 55\,379 } & \multirow{3}{*} { 40.0\% }& \multirow{3}{*} { \citet{2mass}} \\
	$H_\text{2MASS}$ & 16513& & & \\
	$K_{s, \text{2MASS}}$ & 21656 & & & & & \\

	\hline
	\multirow{4}{*} {\begin{tabular}{l} $W_1$ \\ $W_2$ \\ $W_3$ \end{tabular}} &
	\multirow{4}{*} {\begin{tabular}{l} $33792$ \\ $46293$ \\ $123338$ \end{tabular}} &
	\multirow{4}{*} { WISE } & \multirow{4}{*} {
		\begin{tabular}{l}
		$\mathrm{ext\_flg} = 0$\\$\mathrm{cc\_flags} = "000\*"$\\$\mathrm{ph\_qual} = \mathrm{A\ or\ B}$\\ $\delta W_3 / W_3 < 0.05$ \\
		\end{tabular}
	} &\multirow{4}{*} { 26\,182 } & \multirow{4}{*} { 18.9\% }& \multirow{4}{*} {\citet{Wright:2010aa}} \\
	& & & & & & \\
	& & & & & & \\
	& & & & & & \\

	\hline
	All & & & & 15\,765 & 11.4\% & \\
	\hline

	\end{tabular}
	\caption{The sources of the photometric data used for the SED fitting described in \autoref{sub:photometric_fitting}. To aid in comparison to plots, each filter is listed with its isophotal effective wavelength $\lambda_\text{iso}$, criteria used to select the photometry, and original source. The bands are grouped into photometric systems and in order of ascending $\lambda_\text{iso}$. The $\lambda_\text{iso}$ is determined by integrating $f_\lambda = 1$ across the filter, unless directly quoted in the source paper. Due to the poor quality of the photometry, the W4 band has been omitted from the fitting. 
	\newline \noindent
	$^\alpha$ Passbands available from \href{https://www.cosmos.esa.int/web/gaia/iow_20180316}{https://www.cosmos.esa.int/web/gaia/iow\_20180316}. 
	\newline \noindent
	$^\beta$ As this paper was nearing completion \citet{Maiz-Apellaniz:2018aa} was published which presented further revisions to the Gaia system responses. 
	After careful analysis we found that the differences between the fits resulting from grids consisting of both sets of response curves was contained within the majority of formal uncertainties of our resulting stellar properties. 
	}
	\label{tab:data-sources}
\end{table*}

%% file: tab/column-descriptions.tex
\begin{table*}
	\centering
	\input{tab/column-descriptions-tabular}
	\caption{The column descriptions for our output catalogue. }
	\label{tab:column-descriptions}
\end{table*}

%% file: tab/column-descriptions-tabular.tex
\begin{tabular}{| l | l | l |}\hline
Column Name & FITS Name & Column Description \\ 
\hline
Target ID & id & Internal identifier for the object within our catalogue.  \\ 
$\alpha$ & ra & Right ascension of the source as given in the Gaia DR2 catalogue.  \\ 
$\delta$ & dec & Declination of the source as given in the Gaia DR2 catalogue.  \\ 
Gaia Source ID & source\_id & Unique source indentifier of the star within Gaia DR2.  \\ 
$G_{\rm G}$ & phot\_g\_mean\_mag & G-band mean magnitude of the star in the Gaia DR2 catalogue.  \\ 
$G_{\rm BP}$ & phot\_bp\_mean\_mag & Integrated BP mean magnitude of the star in the Gaia DR2 catalogue.  \\ 
$G_{\rm RP}$ & phot\_rp\_mean\_mag & Integrated RP mean magnitude of the star in the Gaia DR2 catalogue.  \\ 
$\delta G_{\rm BP}$ & phot\_bp\_mean\_mag\_err & Uncertainty of the BP mean magnitude of the star in the Gaia DR2 catalogue.  \\ 
$\delta G_{\rm RP}$ & phot\_rp\_mean\_mag\_err & Uncertainty of the RP mean magnitude of the star in the Gaia DR2 catalogue.  \\ 
$F_{\rm BP}$ & phot\_bp\_mean\_flux & Integrated BP mean flux of the star in the Gaia DR2 catalogue.  \\ 
$F_{\rm RP}$ & phot\_rp\_mean\_flux & Integrated RP mean flux of the star in the Gaia DR2 catalogue.  \\ 
$F_{\rm G}$ & phot\_g\_mean\_flux & G-band mean flux of the star in the Gaia DR2 catalogue.  \\ 
$\chi^2_{\rm al}$ & astrometric\_chi2\_al & Astrometric goodness of fit in the Gaia DR2 catalogue.  \\ 
$N_{\rm al, good}$ & astrometric\_n\_good\_obs\_al & No. CCD transits not downweighted in the astrometric solution in the Gaia DR2 Catalogue.  \\ 
$J$ & Jmag\_cds & $J$ band magnitude of the star in 2MASS.  \\ 
$\delta J$ & e\_Jmag\_cds & Uncertainty in $J$ band magnitude of the star in 2MASS.  \\ 
$H$ & Hmag\_cds & $H$ band magnitude of the star in 2MASS.  \\ 
$\delta H$ & e\_Hmag\_cds & Uncertainty in $H$ band magnitude of the star in 2MASS.  \\ 
$K_S$ & Kmag\_cds & $K_s$ band magnitude of the star in 2MASS.  \\ 
$\delta K_s$ & e\_Kmag\_cds & Uncertainty in $K_s$ band magnitude of the star in 2MASS.  \\ 
$W_1$ & W1mag\_cds & $W_1$ band magnitude of the star in AllWISE.  \\ 
$\delta W_1$ & e\_W1mag\_cds & Uncertainty in $W_1$ band magnitude of the star in AllWISE.  \\ 
$W_2$ & W2mag\_cds & $W_2$ band magnitude of the star in AllWISE.  \\ 
$\delta W_2$ & e\_W2mag\_cds & Uncertainty in $W_2$ band magnitude of the star in AllWISE.  \\ 
$W_3$ & W3mag\_cds & $W_3$ band magnitude of the star in AllWISE.  \\ 
$\delta W_3$ & e\_W3mag\_cds & Uncertainty in $W_3$ band magnitude of the star in AllWISE.  \\ 
Model No. & model\_no & Number of the best fitting model in our grid.  \\ 
$5 \log(R / d)$ & norm & Magnitude normalisation calculated by the code to scale $Z_i$ to $m_{i, \text{syn}}$.  \\ 
$\chi^2$ & chisq & Goodness of fit for the best fitting model.  \\ 
$\chi^2_\nu$ & reduced\_chisq & Goodness of fit per degree of freedom for the best fitting model.  \\ 
$P$ & prob & The corresponding probability of being the best fitting model.  \\ 
$R^2 / d^2$ & rsqr\_on\_dsqr & Measured radius squared on distance squared of the target.  \\ 
$R$ & radius & Our measured radius of the target (in units of $R_{\odot}$).  \\ 
$d$ & dist & Estimated distance (in pc) to the target.  \\ 
$d_-$ & dist\_lbound & Lower bound on the confidence interval of the estimated distance (in pc).  \\ 
$d_+$ & dist\_ubound & Upper bound on the confidence interval of the estimated distance (in pc).  \\ 
$N$ & nphot & The number of photometric data points used for the SED fitting.  \\ 
$T_{\rm SED}$ & Teff & Our measured SED temperature of the target.  \\ 
$\log(g)$ & Log(g) & Our measured surface gravity of the target.  \\ 
Photometric Contamination Flag & bad\_phot & Flag showing whether source is consistent with contaminated photometry in Gaia DR2.  \\ 
Astrometric Contamination Flag & bad\_astr & Flag indicating whether source is consistent with contaminated astrometry in Gaia DR2.  \\ 
$T_{\rm SED}$ Edge Flag & bad\_teff & Flag indicating that the measured $T_{\rm SED}$ is at a grid boundary.  \\ 
Good Flag & good & True if none of \textbf{bad\_phot}, \textbf{bad\_astr} or \textbf{bad\_teff} are true for the source.    \\ 
$T_{{\rm SED}-}$ & teff\_lbound & Lower bound on the confidence interval of our measured $T_{\rm SED}$.  \\ 
$T_{{\rm SED}+}$ & teff\_ubound & Upper bound on the confidence interval of our measured $T_{\rm SED}$.  \\ 
$\log(g)_-$ & logg\_lbound & Lower bound on the confidence interval of our measured $\log(g)$.  \\ 
$\log(g)_+$ & logg\_ubound & Upper bound on the confidence interval of our measured $\log(g)$.  \\ 
$R_-$ & radius\_lbound & Lower bound on the confidence interval of our measured radius.  \\ 
$R_+$ & radius\_ubound & Upper bound on the confidence interval of our measured radius.  \\ 
$\chi^2$ 68\% Confidence Contour & tau\_sq\_68\_contour & Value of $\chi^2$ within which 68\% of the confidence is bounded within the cube.  \\ 
Uncertainty Sample & uncer\_sample & Sample for which the uncertainties correspond. For this publication, it will be 'random'.  \\ 
$[ R\ -\ R_{\rm D08}(L) ] / R$ & d08\_inflation & Radius inflation (in \%) from the \citet{Dotter:2008aa} 4Gyr solar metallicity isochrone.  \\ 
$R - R_{\rm fit}(L_{\rm SED})$ & d08\_lum\_corr\_radius & Radius residual when corrected for our luminosity - radius relation.  \\ 
$L_{\rm SED}$ & lum & Luminosity measured for the source when correcting the radius for metallicity.  \\ 
Metallicity Corrected $R$ & corrected\_radius & Radius of the star when corrected for the effects of metallicity.  \\ 
Metallicity Corrected $R_-$ & corrected\_radius\_lbound & Lower bound on the confidence interval of our metallicity corrected radius. \\ 
Metallicity Corrected $R_+$ & corrected\_radius\_ubound & Upper bound on the confidence interval of our metallicity corrected radius. \\ 
$[{\rm Fe} / {\rm H}]$ & fe\_h & Measured metallicity of the target.  \\ 
$[{\rm Fe} / {\rm H}]$ Source & fe\_h\_bibcode & The bibcode corresponding to the source of the metallicity measurement.  \\ 
$F(L_{\rm SED}) [{\rm Fe} / {\rm H}]$ & rad\_corr & The correction applied to our measured radius to correct for metallicity, measured in $R_\odot$.  \\ 
\hline 
\end{tabular}

%% file: tab/tsed-radius-relation-table.tex
\begin{table}
	\centering
	\input{tab/tsed-radius-relation-tabular}
	\caption{A tabulated form of the $T_{\rm SED} - R$ relation presented in \autoref{eq:r-teff-relation} with bounds from \autoref{eq:lower-bound-radii} and \autoref{eq:upper-bound-radii}. The full table is available in the online electronic supplementary material. }
	\label{tab:tsed-radius-relation}
\end{table}

%% file: tab/tsed-radius-relation-tabular.tex
\begin{tabular}{ l l l l }
\hline
$T_{\rm SED}\ ({\rm K})$&$R_{\rm fit}(T_{\rm SED})\ (R_\odot)$&$R_{\rm low} (T_{\rm SED})$&$R_{\rm high} (T_{\rm SED})$\\\hline
3000 &   0.198 &   0.152 &   0.283\\ 
3001 &   0.199 &   0.153 &   0.284\\ 
3002 &   0.199 &   0.154 &   0.285\\ 
3003 &   0.200 &   0.154 &   0.285\\ 
3004 &   0.201 &   0.155 &   0.286\\ 
\hline
\end{tabular}

%% file: tab/lsed-radius-relation-table.tex
\begin{table}
    \centering
	\input{tab/lsed-radius-relation-tabular}
	\caption{A tabulated form of the $L_{\rm SED} - R$ relationship presented in \autoref{eq:dotter-rad-lum-correction} as the sum of the \citet{Dotter:2008aa} isochronal radius with a correction $\Delta R$. We also include the bounds from \autoref{eq:lower-bound-radii-lum-dotter} and \autoref{eq:upper-bound-radii-lum-dotter} and the applied correction. The full table is available in the online electronic supplementary material.}
	\label{tab:lsed-radius-relation}
\end{table}

%% file: tab/lsed-radius-relation-tabular.tex
\begin{tabular}{ l l l l l l l }
\hline
$L_{\rm SED}\ (L_\odot)$&$\Delta R$&$\Delta R_{\rm low}$&$\Delta R_{\rm high}$&$R_{\rm fit}\ (R_\odot)$&$R_{\rm low}$&$R_{\rm high}$\\\hline
  0.003 &   0.017 &   0.004 &   0.031 &   0.195 &   0.182 &   0.209\\ 
  0.004 &   0.018 &   0.004 &   0.032 &   0.219 &   0.206 &   0.233\\ 
  0.005 &   0.018 &   0.004 &   0.033 &   0.240 &   0.226 &   0.254\\ 
  0.006 &   0.019 &   0.005 &   0.033 &   0.258 &   0.244 &   0.273\\ 
  0.007 &   0.019 &   0.005 &   0.034 &   0.275 &   0.260 &   0.289\\ 
\hline
\end{tabular}

%% file: tab/lum-feh-corr.tex
\begin{table}
	\centering
	\input{tab/lum-feh-corr-tabular}
	\caption{The tabulated values for $F(L_{\rm SED})$ in \autoref{eq:rad-feh-corr}. In between $L_{\rm SED}$ points we linearly interpolate neighbouring values. This table is also available in the online electronic supplementary material. }
	\label{tab:lum-feh-corr-grad}
\end{table}

%% file: tab/lum-feh-corr-tabular.tex
\begin{tabular}{| l l | l l | l l |}
\hline
$L_{\rm SED}\ (L_\odot)$ & $F(L_{\rm SED})$ & $L_{\rm SED}$ & $F(L_{\rm SED})$ & $L_{\rm SED}$ & $F(L_{\rm SED})$ \\\hline
0.0035 & -0.0159 & 0.0365 & -0.0744 & 0.0695 & -0.0586\\ 
0.0045 & -0.0192 & 0.0375 & -0.0836 & 0.0705 & -0.0467\\ 
0.0055 & -0.0270 & 0.0385 & -0.0799 & 0.0715 & -0.0534\\ 
0.0065 & -0.0334 & 0.0395 & -0.0645 & 0.0725 & -0.0509\\ 
0.0075 & -0.0342 & 0.0405 & -0.0819 & 0.0735 & -0.0582\\ 
0.0085 & -0.0415 & 0.0415 & -0.0717 & 0.0745 & -0.0369\\ 
0.0095 & -0.0443 & 0.0425 & -0.0784 & 0.0755 & -0.0446\\ 
0.0105 & -0.0414 & 0.0435 & -0.0792 & 0.0765 & -0.0647\\ 
0.0115 & -0.0470 & 0.0445 & -0.0636 & 0.0775 & -0.0448\\ 
0.0125 & -0.0573 & 0.0455 & -0.0820 & 0.0785 & -0.0494\\ 
0.0135 & -0.0497 & 0.0465 & -0.0671 & 0.0795 & -0.0447\\ 
0.0145 & -0.0557 & 0.0475 & -0.0887 & 0.0805 & -0.0670\\ 
0.0155 & -0.0577 & 0.0485 & -0.0674 & 0.0815 & -0.0607\\ 
0.0165 & -0.0663 & 0.0495 & -0.0743 & 0.0825 & -0.0096\\ 
0.0175 & -0.0623 & 0.0505 & -0.0751 & 0.0835 & -0.0447\\ 
0.0185 & -0.0626 & 0.0515 & -0.0644 & 0.0845 & -0.0583\\ 
0.0195 & -0.0668 & 0.0525 & -0.0639 & 0.0855 & -0.0608\\ 
0.0205 & -0.0684 & 0.0535 & -0.0706 & 0.0865 & -0.0323\\ 
0.0215 & -0.0697 & 0.0545 & -0.0784 & 0.0875 & -0.0614\\ 
0.0225 & -0.0744 & 0.0555 & -0.0466 & 0.0885 & -0.0350\\ 
0.0235 & -0.0702 & 0.0565 & -0.0672 & 0.0895 & -0.0434\\ 
0.0245 & -0.0758 & 0.0575 & -0.0552 & 0.0905 & -0.0373\\ 
0.0255 & -0.0789 & 0.0585 & -0.0724 & 0.0915 & -0.0600\\ 
0.0265 & -0.0759 & 0.0595 & -0.0540 & 0.0925 & -0.0463\\ 
0.0275 & -0.0698 & 0.0605 & -0.0846 & 0.0935 & -0.0467\\ 
0.0285 & -0.0803 & 0.0615 & -0.0533 & 0.0945 & -0.0538\\ 
0.0295 & -0.0772 & 0.0625 & -0.0613 & 0.0955 & -0.0351\\ 
0.0305 & -0.0753 & 0.0635 & -0.0601 & 0.0965 & -0.0425\\ 
0.0315 & -0.0830 & 0.0645 & -0.0611 & 0.0975 & -0.0247\\ 
0.0325 & -0.0790 & 0.0655 & -0.0384 & 0.0985 & -0.0279\\ 
0.0335 & -0.0774 & 0.0665 & -0.0544 & 0.0995 & -0.0340\\ 
0.0345 & -0.0818 & 0.0675 & -0.0766 &  & \\ 
0.0355 & -0.0696 & 0.0685 & -0.0497 &  & \\ 
\hline
\end{tabular}